\documentclass[aps,prd,superscriptaddress,nofootinbib,amsmath,amsfonts, preprintnumbers, groupedaddress, showpacs,
10pt,english]{revtex4-1}
\usepackage{amsmath}
\usepackage{amssymb}
\usepackage{babel}
\usepackage{wrapfig}
\usepackage{cancel}

\newcommand{\e}{\mathrm{e}}
\usepackage{relsize,exscale}
\makeatletter

\usepackage{array,multirow,graphicx}
\usepackage{dcolumn}
\usepackage{newlfont}
\usepackage{bm}
\usepackage[colorlinks,citecolor=blue,urlcolor=blue,linkcolor=blue]{hyperref}
\usepackage[figtopcap]{subfigure}
\usepackage{color}

\begin{document}

\allowdisplaybreaks[4]
\tolerance=5000

\date{\today}
\title{Slow-rotating charged black hole solution in dynamical Chern-Simons modified gravity}

\author{G.~G.~L.~Nashed}
\email{nashed@bue.edu.eg}
\affiliation {Centre for Theoretical Physics, The British University, P.O. Box
43, El Sherouk City, Cairo 11837, Egypt}
\author{Shin'ichi~Nojiri}
\email{nojiri@gravity.phys.nagoya-u.ac.jp}
\affiliation{Department of Physics, Nagoya University, Nagoya 464-8602,
Japan \\
\& \\
Kobayashi-Maskawa Institute for the Origin of Particles and the Universe,
Nagoya University, Nagoya 464-8602, Japan }

\begin{abstract}
The Chern-Simons (CS) gravity is a modified theory of Einstein's general relativity (GR).
The CS theory arises from the low energy limit of string theory which involves anomaly correction to the Einstein-Hilbert action.
The CS term is given by the product of the Pontryagin density with a scalar field.
In this study, we derive a charged slowly rotating black hole (BH) solution.
The main incentives of this BH solution are axisymmetric and stationary and form distortion of the Kerr-Newman BH solution with a dipole scalar field.
Additionally, we investigate the asymptotic correction of the metric with the inverse seventh power of the radial distance to the BH solution,
This indicates that it will escape any meaningful constraints from weak-field experiments.
To find this kind of BHs by observations, we investigate the propagation of the photon near the BH and we show that the difference between the
left-rotated polarization and the right-handed one could be observed as stronger than the case of the Kerr-Newman BH.
Finally, we derived the stability condition using the geodesic deviations.
\end{abstract}


\maketitle

\section{Introduction}\label{intro}

Is Einstein's theory of general relativity (GR) still right?
The solar system~\cite{Will:2014kxa, Bertotti:2003rm} and the binary pulsar observations~\cite{Stairs:2003eg} have affirmed
that GR could be right to ultra-precision in the weak field domain, where the gravitational field is weak and non-dynamical.
The new gravitational wave observations detected by the advanced Laser Interferometer Gravitational wave Observatory
(LIGO)~\cite{Abramovici:1992ah,LIGOScientific:2007fwp,Harry:2010zz,LIGOScientific:2007fwp} and
the advanced Virgo detector~\cite{Caron:1997hu, Giazotto:1988gw, VIRGO:2014yos} have ensured that GR appears to be on a solid base,
even in the rather strong gravity region, where the gravitational interaction is highly dynamical~\cite{Berti:2018cxi, Berti:2018vdi}.

According to these observations, one might wonder if additional tests of GR are still urgent.
The main motivations for continuous checking are observational and theoretical.
According to the observation viewpoint, the late-time acceleration of the universe as inferred through supernova~\cite{SupernovaSearchTeam:2004lze},
the rotation curves of galaxies~\cite{Rubin:1970zza}, and different other observations~\cite{Helbig:1999cu,Clowe:2006eq,SDSS:2006lmn,Planck:2018vyg}
point to anomalies that might be either solved by mentioning the presence of dark energy~\cite{Huterer:1998qv} and dark matter~\cite{Carr:1994ci},
otherwise, by amending GR on large scales~\cite{Scherrer:2004au}.
According to a theoretical viewpoint, the inherent inconsistency of GR with quantum mechanics has given rise to the emergence of amendment theories
that try a reconciliation~\cite{Ashtekar:2004eh, Rovelli:2004tv}.
In any case of the argumentation, additional testing could supply extra hints that might help in solving some of these oddities.

The gravitational parity violation as evolved in the dynamical Chern-Simons (dCS) gravity is the specific modification of GR
that has attracted some interests~\cite{Jackiw:2003pm, Alexander:2009tp}.
This theory amends the Einstein GR by a dynamical (pseudo) scalar field that couples non-minimally with curvature by the Pontryagin density.
The magnitude of deviation in dCS theory from GR is ruled by the size of its dimensional coupling parameter $\xi$.
Nowadays, dCS gravity theory is understood as an effective model, which is valid in a region where the energy scale or curvature is small compared
with some cut-off scale, i.e., when the coupling $\xi$ is small.
The reason is the fact that the dCS theory is motivated by heterotic string theory at 4-dimensional compactification
and a low-curvature expansion~\cite{Alexander:2004xd}, from effective field theories of inflation~\cite{Weinberg:2008hq},
and loop quantum gravity at the promotion of the Barbero-Immirzi parameter to the field in the existence of matter~\cite{Taveras:2008yf, Calcagni:2009xz}.

The gravitational parity violation in dCS theory appears in the systems that break parity by the presence of a preferred axis like the one prescribed
by angular momentum in the dynamical system~\cite{Yunes:2008ua, Alexander:2007zg, Smith:2007jm, Alexander:2007vt, Nakamura:2018yaw, Yagi:2013mbt}.
An example is the isolated spinning BH, whose solutions in the dCS theory are known up to the fifth order
in the slow-rotation expansion~\cite{Yunes:2009hc, Yagi:2012ya, Maselli:2017kic}, and in the near-maximum~\cite{McNees:2015srl, Chen:2018jed}.
Another example is the spinning planet or star, whose behaviors in the solar system have been investigated and observed for approximate horn.
Therefore, we expect that any dCS deformation from GR forecasts in the spin dynamics of the solar system may be employed to limit the theory.

The CS theory is classified into the non-dynamical type and the dynamical type.
In the dynamical case, the (pseudo) scalar field evolves in time by following the field equations.
In the non-dynamical type, there is no kinetic term for the scalar field and therefore the variation with respect to the scalar field gives the constraint that
the gravitational Pontryagin density $R\tilde R$ identically vanishes although the Chern-Simons term gives non-trivial-contributions.
The two types present amended equations of motion compared to general relativity field equations.

In the non-dynamical CS model, Alexander and Yunes~\cite{Alexander:2007vt} have shown that the gravitomagnetic sector of the metric is amended
in the solar system, yielding a new parameterized post-Newtonian parameter~\cite{{Alexander:2007zg}}.
Similarly, Smith, et~al.~\cite{Smith:2007jm} have calculated the correction for the spin-precession using a uniform distribution of density objects
which uniform rotation and compared their result to the observations using the Gravity Probe B (GPB) experiment~\cite{Everitt:2011hp}
and the LAGEOS satellites to put a constraint on the non-dynamical theory.
Soon, Ali-Ha{\"i}moud and Chen~\cite{Ali-Haimoud:2011zme} studied the dynamical theory and computed the extra correction to the gravitomagnetic of the metric
and put an approximate constraint on the theory of $\xi_\mathrm{CS}^{1/4} \lesssim \mathcal{O} \left( 10^8\, \mathrm{km} \right)$.
Moreover, the possibility of comparable constraints using quantum/Sagnac interferometry has been investigated in~\cite{Okawara:2012bi, Okawara:2013wc, Kikuchi:2014mva}.
Soon, observing gravitational waves emitted by the rotating BHs will provide us with an eight-degree improvement in such constraints~\cite{Loutrel:2018rxs},
once these discoveries are made that are powerful enough to break degeneracies between the spins of the objects and the dCS deformation.
{ Rotating black hole solutions in the $(3+1)$-dimensional Chern-Simons modified gravity theory are investigated
by taking account of perturbation around the Schwarzschild solution \cite{Konno:2007ze}.
A detailed study of the BH solutions in the Chern-Simons gravity has revealed that at least two different limits of the Kerr BH are solutions
to the modified field equations \cite{Grumiller:2007rv}.
In the extended Chern-Simons modified gravity, an investigation of a rotating BH has been carried out \cite{Konno:2009kg}.
A study of the null geodesics corresponding to a slowly rotating BH in Chern Simons gravity with a small coupling constant revealed
that the photon orbits are separable as in the Kerr geometry \cite{Amarilla:2010zq}.
A study of the null geodesics corresponding to a slowly rotating BH in Chern Simons gravity, with a small coupling constant has shown
that the photon orbits are separable as in the Kerr geometry \cite{Brihaye:2016lsx}.
Four-dimensional homogeneous static and rotating black strings, with and without torsion in the dCS modified gravity, are presented \cite{Cisterna:2018jsx}.}

Yunes and Pretorius \cite{Yunes:2007ss} have derived a non-trivial uncharged black solution in the dCS theory that generates the parity violation.
The purpose of the present study is to derive a new weakly charged rotating BH solution in the framework of the dCS gravitational theory.

The construction of this study is as follows:
In Sec.~\ref{ABC}, we present the main tools of CS-modified gravity.
In Sec.~\ref{axisym}, we solve the field equations of the dCS gravity for the line element prescribing slowly rotating BH which is valid for small
CS coupling constants.
In Sec.~\ref{properties}, we consider some of the related physics to the derived solutions to understand its physical properties.
{ In Sec.~\ref{geod}, we study the motion of the CS BH presented in this study and derive the conserved quantities of this BH and its orbital period. }
In Sec.~\ref{polarization}, we investigate the propagation of the photon near the BH and we consider the difference of the polarization from that of the Kerr-Newman BH.
In Sec.~\ref{conclusions}, we conclude the main results of this study and present possible future work.

The following conventions are used throughout the present study:
We use four-dimensional space-times that have the following signature
$(-,+,+,+)$~\cite{Misner:1973prb}, square brackets and parentheses mean
anti-symmetrization and symmetrization, respectively, i.e., $T_{[\mu\nu]}=\frac12 (T_{\mu\nu}-T_{\nu\mu})$ and $T_{(\mu\nu)}=\frac12
(T_{\mu\nu}+T_{\nu\mu})$. The partial
derivatives are refereed by commas (e.g.~$\partial
\varphi/\partial r=\partial_r\varphi=\varphi_{,r}$).
The Einstein summation $A_\mu B^\mu = \sum_{\mu=0,1,2,3} A_\mu B^\mu$ is applied and geometrized units $G=c=1$ are used.

\section{CS modified gravitational theory}\label{ABC}

In this section, we will prescribe the subjects which give a full construction of the CS-modified gravitational theory and present some notation
\cite{Alexander:2009tp}.

\subsection{ABC of CS gravitational theory}

The action of the CS gravitational theory is given as,
\begin{align}
\label{CSaction}
S = S_1 + S_2 + S_3+S_4+S_5\,,
\end{align}
where $S_1$ is the Einstein Hilbert action defined by,
\begin{align}
\label{EH-action}
S_1 = \frac{1}{2\kappa^2} \int_V d^4x \sqrt{-g} , ,
\end{align}
$S_2$ is the Chern-Simon action given by,
\begin{align}
\label{CS-action}
S_2= \frac\sigma{4} \int_V d^4x \sqrt{-g} \varphi R \tilde R \,,
\end{align}
$S_3$ is the action of the (pseudo) scalar field $\varphi$ with the potential $U(\varphi)$ defined by,
\begin{align}
\label{Theta-action}
S_3 = - \frac{\sigma_1}{2} \int_V d^4x \sqrt{-g} \left[ g^{\mu\nu}
\left(\nabla_\mu \varphi\right) \left(\nabla_\nu \varphi\right) + 2 U(\varphi) \right]\, ,
\end{align}
$S_4$ is the action of the electromagnetic field given by,
\begin{align}
\label{EMaction}
S_4= \frac{\sigma_1}{4} \int_V d^4x\sqrt{-g} F^2 \quad \mathrm{where} \quad F^2=F_{\mu \nu}F^{\mu\nu}\,,
\quad \mbox{and} \quad F_{\mu \nu} =A_{\mu, \nu} - A_{\nu, \mu}\,,
\end{align}
and $S_5$ is the action of the matters defined by,
\begin{align}
\label{matteraction}
S_5= \int_V d^4x \sqrt{-g} {L}_\mathrm{mat}.
\end{align}
Here $ V$ is the bulk region in the space-time manifold.

The following notations are used throughout the whole of the present study:
$\sigma$ and $\sigma_1$ are dimensional constants, $\kappa^2 = 8 \pi G$, $\nabla_{\rho}$ is the covariant derivative,
$R$ is the Ricci scalar, and $g$ is the determinant of the metric.
The term\\ $R \tilde R $ is the Pontryagin density defined by,
\begin{align}
\label{pontryagindef}
 R \tilde R = R^\nu_{\ \mu\rho\sigma} \tilde R^{\mu\ \rho\sigma}_{\ \nu}\,,
\end{align}
where $\tilde R^{\mu\ \rho\sigma}_{\ \nu}$ is the dual Riemann tensor which is defined by,
\begin{align}
\label{Rdual}
\tilde R^{\mu\ \rho\sigma}_{\ \nu}=\frac{1}{2} \epsilon^{\rho\sigma\tau\eta}R^\mu_{\ \nu\tau\eta}\,,
\end{align}
where $\epsilon^{\rho\sigma\tau\eta}$ is the 4-dimensional Levi-Civita tensor which is a totally skew-symmetric tensor with $\epsilon^{0123}=-1$.

In this study, the CS scalar field $\varphi$ is a function of the space-time coordinates that parameterize the deviation from the Einstein GR.
If $\varphi = \mathrm{constant}$, the CS gravitational theory will be equivalent to the Einstein GR theory because the Pontryagin density is
the total divergence of the CS topological current $K^\mu$ which is given by,
\begin{align}
\nabla_\mu K^\mu = \frac{1}{2} R \tilde R\, ,
\label{eq:curr1}
\end{align}
where
\begin{align}
K^\mu =\epsilon^{\mu\nu\rho\sigma} \Gamma^{\tau}_{\nu\eta} \left(\partial_\rho\Gamma^\eta_{\sigma\tau}
+\frac{2}{3} \Gamma^\eta_{\rho\xi}\Gamma^\xi_{\sigma\tau}\right)\,,
\label{eq:curr2}
\end{align}
with $\Gamma^\mu_{\nu\rho}$ being the Christoffel connection.
Eq.~\eqref{eq:curr1} tells that $S_2$ can be rewritten in the form \cite{Yunes:2007ss}:
\begin{align}
\label{CS-action-K}
S_2 = \frac\sigma{2} \int_{\partial V} dS_\mu \varphi K^\mu - \frac\sigma{2} \int_V d^4x \sqrt{-g}
\left(\nabla_\mu \varphi \right) K^\mu\,.
\end{align}
Here $\partial V$ is the boundary hypersurface of the space-time manifold and $dS_\mu$ is the infinitesimal area of the hypersurface $\partial V$.
We often neglect the first term in Eq.~\eqref{CS-action-K} because it is irrelevant to the variation in the bulk space-time manifold when we derive the field equations.

The variations of the action \eqref{CSaction} w.r.t. the metric, the scalar field, and the electromagnetic field give the following field equations,
\begin{align}
\label{eom}
R_{\mu\nu} + 2\kappa^2 \sigma_1 C_{\mu\nu} =&\, \kappa^2 \left(T_{\mu\nu} - \frac{1}{2} g_{\mu\nu} T \right)\, , \\
\label{eq:constraint}
\sigma_1 \square \varphi =&\, \sigma_1 \frac{dV}{d\varphi} - \frac\sigma{4} R \tilde R \,, \\
\label{eq:elct}
0=&\, \partial_\nu \left( \sqrt{-g} F^{\mu\nu} \right)\,,
\end{align}
where $R_{\mu\nu}$ is the Ricci tensor and $\square = \nabla_\mu \nabla^\mu$ is the D'Alembertian operator.
The term $C_{\mu\nu}$ is the C-tensor which is defined by, {
\begin{align}
\label{Ctensor}
C^{\mu\nu} = v_\rho
\epsilon^{\rho\sigma \tau (\mu}\nabla_\tau R^{\nu)}_{\ \sigma} +v_{\rho\sigma} \tilde R^{\sigma(\mu\nu)\rho}\,,
\end{align}
}
with
\begin{align}
\label{v}
v_\mu=\nabla_\mu\varphi\,,\qquad \mathrm {and } \qquad
v_{\mu\nu}=\nabla_\mu \nabla_\nu \varphi\,.
\end{align}
As a final point, the total stress-energy tensor is written as,
\begin{align}\label{Tab-total}
T_{\mu\nu} = T^\mathrm{mat}_{\mu\nu} + T_{\mu\nu}^\varphi+ T_{\mu\nu}^\mathrm{EM},
\end{align}
with $T^\mathrm{mat}_{\mu\nu}$ being the matter stress-energy tensor (which we will neglect in the present study),
 $T_{\mu\nu}^\mathrm{EM}$ and $T_{\mu\nu}^\varphi$, are the stress-energy tensors of the electromagnetic and scalar fields figured as,
\begin{align}
\label{Tab-theta}
T_{\mu\nu}^\mathrm{EM} =&\, \frac{\sigma_1}{4\pi}\left[ g_{\rho\sigma} F_\nu^{\ \rho} F_\mu^{\ \sigma}
 - \frac{1}{4} g_{\mu\nu} F^2\right]\,,\nonumber\\
T_{\mu\nu}^\varphi =&\, \sigma_1 \left[ \left(\nabla_\mu \varphi\right) \left(\nabla_\nu \varphi\right)
 - \frac{1}{2} g_{\mu\nu}\left(\nabla_\rho \varphi\right) \left(\nabla^\rho \varphi\right)
 - g_{\mu\nu} U(\varphi) \right]\,.
\end{align}
In the realm of the CS gravitational theory, the strong equivalence principle, i.e.,
$\left( \nabla^\nu T_{\mu\nu}^\mathrm{mat} = 0\right)$, is verified provided that the equation of motions
of the scalar field $\varphi$, Eq.~\eqref{eq:constraint} is satisfied.
This is because if we consider the derivative of Eq.~\eqref{eom}, the first term on the l.h.s. vanishes due to the Bianchi identities,
while the second term is proportional to the Pontryagin density through the form,
\begin{align}
\label{nablaC}
\nabla_\mu C^{\mu\nu} = - \frac{1}{8} v^\nu R \tilde R .
\end{align}
The verification of Eq.~\eqref{nablaC} yields Eq.~\eqref{eq:constraint}.

To finish this section, we are going to discuss the dimensions of the coupling constants used in this study and the scalar field $\varphi$ and
the electromagnetic field $A_\mu$.
By fixing the dimensions of $(\sigma,\sigma_1,\varphi, A_\mu)$, the units of the other constants are also fixed.
For example, if the CS scalar field and the electromagnetic field have the dimensions $[\varphi] = [A_\mu] = l^s$,
then $[\sigma] = l^{2 - s}$ and $[\sigma_1] = l^{-2s}$, where $l$ expresses the dimension of length.
The CS scalar $\varphi$ and the electromagnetic field $A_\mu$ is often dimensionless, which requires $[\sigma] = \left[ A_\mu \right] = l^2$
and $\sigma_1$ be dimensionless\footnote{In this study, we use the geometrical units with $G = c = 1$,
and thus, the action has the units of $l^2$.
Therefore, if natural units are used where $\hbar = c = 1$, then the action will be dimensionless and therefore if $[\varphi] = [A_\mu] = l^{s}$
then $[\sigma] = = \left[ A_\mu \right] l^{-s}$ and $[\sigma_1] = l^{-2 s - 2}$.}.
Other selection is to put $\sigma = \sigma_1$, thus putting $S_2$, $S_3$, and $S_4$ on equal footing and we have
$[\varphi] [A_\mu] = l^{-2}$.
We will leave this arbitrariness because the results of the previous studies are based on the different choices of the unit.

\section{Rotating Charged BH solution in dynamical CS gravity }\label{axisym}

Now, we are going to study rotating charged BHs in the dynamical construction of the modified CS theory.
The study of stationary axisymmetric line elements in the frame of the CS gravitational theory
without doing any approximation in the calculation will be a tedious task.
Therefore, we will use a pair of approximations.
Thence, we will proceed to solve the modified CS equation of motions to second order in perturbation expansion.
In the following, we only consider the case that the potential $U(\varphi)$ for the scalar field $\varphi$ vanishes, $U(\varphi)=0$.

\subsection{The process of approximation}
\label{approx}

Now we will use two approximation processes: a slow-rotation and small-coupling approximations.
The small-coupling process deals with the modified CS term as a small distortion of GR,
which allows expanding the metric and the gauge potential $A_\mu$ (up to second order) as follows,
\begin{align}
\label{small-cou-exp0}
g_{\mu\nu} = g_{\mu\nu}^{(0)} + \xi g^{(1)}_{\mu\nu}(\varphi) + \xi^2 g^{(2)}_{\mu\nu}(\varphi)\,, \quad
A_a=A_\mu^{(0)} + \xi A^{(1)}_\mu + \xi^2 A^{(2)}_\mu \,
\end{align}
where $g_{\mu\nu}^{(0)}$ and $A_\mu^{(0)}$ are the background metric and charge which satisfy the Einstein GR field equations, such as the Kerr-Newmann metric,
while $g_{\mu\nu}^{(1)}(\varphi)$, $ A^{(1)}_\mu$, $g_{\mu\nu}^{(2)}(\varphi)$, and $A^{(2)}_\mu $ are the first and the second-order perturbation coming from the CS corrections.
The parameter $\xi$ refers to the order of the small-coupling approximation, which we will define soon.

On the other hand, the slow-rotation approximation allows to re-expand the background and the $\xi$-perturbations in powers
of the Kerr-Newmann rotation parameter $a_\mathrm{KN}$.
Therefore, the background metric and the metric perturbation yield the following form,
\begin{align}
\label{small-cou-exp}
g_{\mu\nu}^{(0)} =&\, \eta_{\mu\nu}^{(0,0)} + \epsilon h_{\mu\nu}^{(1,0)} + \epsilon^2 h_{\mu\nu}^{(2,0)}\, , \nonumber \\
\xi g_{\mu\nu}^{(1)} =&\, \xi h_{\mu\nu}^{(0,1)} + \xi \epsilon h_{\mu\nu}^{(1,1)} + \xi \epsilon^2 h_{\mu\nu}^{(2,1)},
\nonumber \\
\xi^2 g_{\mu\nu}^{(2)} =&\, \xi^2 h_{\mu\nu}^{(0,2)} + \xi^2 \epsilon h_{\mu\nu}^{(1,2)} + \xi^2 \epsilon^2 h_{\mu\nu}^{(2,2)}\, ,\nonumber \\
A_\mu^{(0)} =&\, \eta_\mu^{(0,0)} + \epsilon A_\mu^{(1,0)} + \epsilon^2 A_\mu^{(2,0)}\, , \nonumber \\
\xi A_\mu^{(1)} =&\, \xi A_\mu^{(0,1)} + \xi \epsilon A_\mu^{(1,1)} + \xi \epsilon^2 A_\mu^{(2,1)}\, , \nonumber \\
\xi^2 A_\mu^{(2)} =&\, \xi^2 A_\mu^{(0,2)} + \xi^2 \epsilon A_\mu^{(1,2)} + \xi^2 \epsilon^2 A_\mu^{(2,2)}\, ,
\end{align}
where the parameter $\epsilon$ stands for the order of the slow-rotation expansion, which we will also define soon.
We must remind ourself that the notation $h^{(n,m)}_{\mu\nu}$ labels for terms of $\mathcal{O} \left( n, m \right)$,
which stands for a term of $\mathcal{O} \left(\epsilon^n \right)$ and $\mathcal{O} \left(\xi^m \right)$.
As an example, in Eq.~\eqref{small-cou-exp}, $\eta_{\mu\nu}^{(0,0)}$ and $\eta_\mu^{(0,0)}$ are the background metric when the rotation parameter vanishing, i.e.,
$a_\mathrm{KN} = 0$, whilst $h_{\mu\nu}^{(1,0)}$, $h_{\mu\nu}^{(2,0)}$, $A_\mu^{(1,0)}$ and $A_\mu^{(2,0)}$ are first and second-order perturbations
of the background metric and charge in the spin parameter.

Combine both approximation processes, we obtain a bivariate expansion in terms of two independent parameters $\xi$ and $\epsilon$,
which yields the second perturbation order of the metric and the electromagnetic field in the following forms,
\begin{align}
g_{\mu\nu} =&\, \eta_{\mu\nu}^{(0,0)} + \epsilon h_{\mu\nu}^{(1,0)} + \xi h_{\mu\nu}^{(0,1)} + \epsilon \xi h_{\mu\nu}^{(1,1)} + \epsilon^2 h_{\mu\nu}^{(2,0)} + \xi^2 h_{\mu\nu}^{(0,2)}
\, , \nonumber\\
A_\mu =&\, A_\mu^{(0,0)} + \epsilon A_\mu^{(1,0)} + \xi A_\mu^{(0,1)} + \epsilon \xi A_\mu^{(1,1)} + \epsilon^2 A_\mu^{(2,0)} + \xi^2 A_\mu^{(0,2)}\,.
\end{align}
The first-order expressions, refers to expressions of $\mathcal{O}\left(1,0\right)$ or $\mathcal{O}\left(0,1\right)$,
while second-order terms refers to $\mathcal{O}\left(2,0\right)$, $\mathcal{O}\left(0,2\right)$, or $\mathcal{O}\left(1,1\right)$.

In this study, the slow-rotation process is the expansion of the Kerr-Newmann parameter, $a_\mathrm{KN}$,
and therefore its dimensionless expansion parameter $\epsilon$ should be $\epsilon=a_\mathrm{KN}/M$.

\subsection{The slowly rotating charged BH solution}\label{slow-rot}

The slowly rotating expansion of the background metric can be formulated using the Hartle-Thorne approximation~\cite{Thorne:1984mz, Hartle:1968si},
where the line element can be parameterized as follows,
\begin{align}
\label{slow-rot-ds2}
ds^2 =&\, -h \left[1 + h_1\left(r,\theta\right)\right] dt^2
+ \frac{1}{h} \left[1 + h_2\left(r,\theta\right)\right] dr^2
+ r^2 \left[1 + h_3\left(r,\theta\right) \right] d\theta^2 \nonumber \\
&\, +r^2 \sin^2{\theta} \left[1 + h_4\left(r,\theta\right) \right] \left[ d\phi - \omega\left(r,\theta\right) dt \right]^2\,,
\end{align}
where $h$ is defined as $h = 1 - \frac{2 M} r +\frac{q^2}{r^2}$, which is given by the Reissner-Nordstr\"om solution,
with $M$ being the mass of the charged BH and $q$ being the electric charge in the absence of the CS expression.
In Eq.~\eqref{slow-rot-ds2}, we use the Boyer-Lindquist coordinates, i.e., $(t,r,\theta,\phi)$ and the
perturbations of the metric are $h_1\left(r,\theta\right)$, $h_2\left(r,\theta\right)$, $h_3\left(r,\theta\right)$, $h_4\left(r,\theta\right)$,
and $\omega\left(r,\theta\right)$.

The metric~\eqref{slow-rot-ds2} is rewritten similar to the one presented in~\cite{Thorne:1984mz, Hartle:1968si}, however,
the metric perturbations should be expanded in a series in both $\xi$ and $\epsilon$.
By the second order expansion, we have,
\begin{align}
\label{cons}
h_1\left(r,\theta\right) =&\, \epsilon h_1^{(1,0)} + \epsilon \xi h_1^{(1,1)} + \epsilon^2 h_1^{(2,0)}\, , \nonumber \\
h_2\left(r,\theta\right) =&\, \epsilon h_2^{(1,0)} + \epsilon \xi h_2^{(1,1)} + \epsilon^2 h_2^{(2,0)}\, , \nonumber \\
h_3\left(r,\theta\right) =&\, \epsilon h_3^{(1,0)} + \epsilon \xi h_3^{(1,1)} + \epsilon^2 h_3^{(2,0)}\, , \nonumber \\
h_4\left(r,\theta\right) =&\, \epsilon h_4^{(1,0)} + \epsilon \xi h_4^{(1,1)} + \epsilon^2 h_4^{(2,0)}\, , \nonumber \\
\omega\left(r,\theta\right) =&\, \epsilon \omega^{(1,0)} + \epsilon \xi \omega^{(1,1)} + \epsilon^2 \omega^{(2,0)}\, .
\end{align}
Equations~\eqref{cons} have no expressions of $\mathcal{O}(0,0)$ because those terms are already involved in the Reissner-Nordstr\"om structure of Eq.~\eqref{slow-rot-ds2}.
Also, we assume that when the rotation parameter of Kerr-Newman vanishes, i.e., $a_\mathrm{KN}\rightarrow 0$, we got Reissner-Nordstr\"om space-time as a solution,
which ensures that all terms of $\mathcal{O} \left(0, m \right)$ vanish.
Therefore, the CS expression should be linear in the Kerr-Newmann spin parameter $a_\mathrm{KN}$.
Using the slow-rotation limit of the Kerr-Newman metric in GR, the metric and charge perturbations proportional to $\xi^{0}$ in the first order are given by,
\begin{align}
h_1^{(1,0)} = h_2^{(1,0)} = h_3^{(1,0)} = h_4^{(1,0)} =0\,, \quad
\omega^{(1,0)}=\frac{\left[ 2 Mr-q^2 \right] a_\mathrm{KN}}{r^4}, \quad A_t^{(1,0)}=\frac{q \sin^2\theta a_\mathrm{KN}} r ,
\end{align}
and in the second order,
\begin{align}
h_1^{(2,0)} =&\, \frac{ {a_\mathrm{KN}}^2 \left(2Mr-q^2 \right)}{h r^4} \left( \cos^2{\theta} + \frac{2 Mr-q^2}{r^2} \sin^2{\theta} \right)\,,\quad
h_2^{(2,0)} =\frac{{a_\mathrm{KN}}^2}{r^2} \left( \cos^2{\theta} - \frac{1}{h} \right)\,, \nonumber \\
h_3^{(2,0)} =&\, \frac{{a_\mathrm{KN}}^2}{r^2} \cos^2{\theta}\,, \quad
h_4^{(2,0)} = \frac{{a_\mathrm{KN}}^2}{r^2} \left(1 + \frac{2 Mr-q^2}{r^2} \sin^2{\theta} \right)\,, \quad
A_t^{(2,0)}=-\frac{q{ {a_\mathrm{KN}}^2}}{r^3}\,, \quad \omega^{(2,0)} = 0\,,
\end{align}
whose expressions coincide with those in the Kerr solution when $q=0$ \cite{Yunes:2007ss}.
All the fields are expanded by small-coupling and slow-rotation approximation parameters, including the CS field.
To derive the leading-order corrections for $\varphi$, we must yield to the evolution equation~\eqref{eq:constraint}.
From Eq.~\eqref{eq:constraint}, we obtain $\partial^2 \varphi \sim (\sigma_1/\sigma) R \tilde R $, where
the Pontryagin density equal to zero up to order in $a_\mathrm{KN}/M$.
Thus, the first order behavior of the CS scalar field must be $\varphi \sim \left(\sigma_1/\sigma \right) \left( a_\mathrm{KN}/M \right)$, which is proportional to $\epsilon$.
Additionally, the assumption that the Reissner-Nordstr\"om metric is the unique charged solution with vanishing angular momentum,
we should have $\varphi^{(0,s)} = 0$ for all $s$.
The analysis given in \cite{Yunes:2007ss} to derive a detorsion to the Kerr solution in the dCS construction did not take into account the effect of the charge.
On the other hand, in the present study, we expand the analysis presented in \cite{Yunes:2007ss} to take into account the effect of the charge.
By employing Eq.~\eqref{eq:constraint} to Eq.~\eqref{slow-rot-ds2} and by using Eq.~\eqref{cons}, we obtain,
\begin{align}
\label{th-ansatz}
\varphi = \epsilon \varphi^{(1,0)}\left(r,\theta\right) + \epsilon \xi \; \varphi^{(1,1)}\left(r,\theta\right) + \epsilon^2 \varphi^{(2,0)}\left(r,\theta\right)\,.
\end{align}
Now we are going to apply the process described above to solve the above-modified field equations, by stressing the evolution equation of the dCS scalar.
Up to the order $\mathcal{O}(1,0)$, we obtain the evolution equation in the following form,
\begin{align}
\label{1st-eq1}
h \varphi^{(1,0)}_{,rr} + \frac{2} r \varphi^{(1,0)}_{,r} \left( 1 - \frac M r \right) + \frac{1}{r^2} \varphi^{(1,0)}_{,\theta \theta} + \frac{\cot{\theta}}{r^2} \varphi^{(1,0)}_{,\theta}
= - \frac{24 \sigma M\left(Mr-q^2\right) \left(3Mr-2q^2\right) \cos{\theta}}{\sigma_1 r^9} \frac{a_\mathrm{KN}} M \,.
\end{align}
The solution of partial differential equation \eqref{1st-eq1} is a linear composition of the homogeneous and the particular solution:
$\varphi^{(1,0)} = \varphi^{(1,0)}_\mathrm{Hom} + \varphi^{(1,0)}_\mathrm{Part}$.
The variables of the homogeneous solution for the equation can be separated,
\begin{align}
\label{hom}
\varphi^{(1,0)}_\mathrm{Hom}\left(r,\theta\right) = \varphi(r)\varphi(\theta)\, .
\end{align}
Eq.~\eqref{hom} shows that the partial differential equation becomes a set of ordinary differential equations for $\varphi(r)$ and $\varphi(\theta)$,
whose solutions take the following forms,
\begin{align}
\label{Hom-sol-1}
\varphi(r) =&\, c_1 \mathcal H\left[\left[\frac{s}{2},\frac{s}{2}\right],s,\frac{2 \sqrt H}{r-M+\sqrt H} \right] \left(r-M+\sqrt H\right)^{-\frac{s}{2}}\nonumber\\
&\, + c_{2} \mathcal H\left[\left[\frac{s_1}{2},\frac{s_1}{2}\right],s_1,\frac{2 \sqrt H}{r-M+\sqrt H} \right] \left(M-r+\sqrt H\right)^{-\frac{s_1}{2}}, \quad M>q\,,
\nonumber \\
\varphi(\theta) =&\, c_3 L \left( -\frac{s}{2},\, \cos \theta \right) + c_4 L_1 \left(-\frac{s}{2}, \cos\theta \right) \,,
\end{align}
where $H=M^2-q^2$, and $\mathcal H(\cdots)$ are generalized hypergeometric functions\footnote{{
The generalized hypergeometric function $\mathcal H \left( \left[n_1, n_2,\cdots, n_p \right], \left[d_1, d_2, \cdots, d_q \right], z\right)$ is
generally defined by,
\[
\mathcal H \left( \bm{n}, \bm{d}, z \right) = \sum_{k=0}^\infty \frac{\prod_{i=1}^p \mathrm{PS} \left( n_i, k \right)}
{\prod_{j=1}^q \mathrm{PS} \left( d_j,k \right)} \frac{z^k}{k!}\, ,
\]
where $\bm{n} = \left[ n_1,n_2,\cdots, n_p \right]$, $\bm{d}= \left[ d_1,d_2,\cdots, d_q \right]$ and $\mathrm{PS}(n,k)$
is the Pochhammer symbol, $\mathrm{PS}(n,k) \equiv \prod_{j=0}^{k-1} \left( n + j \right)$.
$H(\cdots)$'s in (\ref{Hom-sol-1}) correspond to $p=2$ and $q=1$.
}},
$L(\cdot)$ is the Legendre polynomial of the first kind\footnote{{
The Legendre polynomial of the first kind is defined by,
\[
L(b,z)= \mathcal H\left( [-b,b+1],[1],\frac{1}{2}(1-z) \right) \, .
\] }},
${L_1}(\cdot)$ is the Legendre polynomial of the second kind\footnote{
The Legendre polynomial of the second kind is defined by,
\[
L_1(b,z)=\frac{\sqrt{\pi} \Gamma(1+b) \mathcal H \left( \left[ 1 + \frac{b}{2}, \frac{1}{2} + \frac{b}{2} \right], \left[ \frac{3}{2}+b \right],\frac{1}{z^2} \right)}
{2 z^{1+b} \Gamma \left( \frac{3}{2}+b \right)2^b}\, .
\] } $c_{i}$, $i=1, \cdots, 4$ are constants of integration, and the constants $s$ and $s_1$ are defined by,
\begin{align}
\label{tilde-alpha}
s= 1 - \sqrt{1 - 4 c_5}\,,
\quad
s_1 = 1 + \sqrt{1 - 4 c_5},
\end{align}
where $c_5$ is the constant of integration that arises through the separation of variables.

We study the solution of $\varphi^{(1,0)}$ in detail to understand the physics in the constants of integration that appear in it.
For this purpose, we will consider the behavior of the solution when $r \gg M$ and obtain,
\begin{align}
\varphi(r) \sim c_1 \left[ 1 + \frac{Mr-q^2}{2 r^2} s \right] r^{-\frac{s}{2}}
+ c_{2} \left[ 1 + \frac{Mr-q^2}{2 r^2} s_1\right] r^{-\frac{s_1}{2}} .
\end{align}
Moreover, we require the scalar field $\varphi$ to have a real value, then the constants $s$ and $s_1$ must be real, $s,\, s_1 \in \Re$,
which requires $c_5 < 1/4$ as we find in (\ref{tilde-alpha}).
Moreover, if we also require the scalar field $\varphi$ to have finite total energy,
thence $\varphi$ must decrease to a constant asymptotically faster than $1/r$, which tells $s > 2$ and $s_1>2$.
The first constraint cannot be satisfied when $c_5 < 1/4$, therefore we find $c_1 = 0$, and the second constraint yields $c_5 < 0$.
Thus, the constraints coming from the requirement of the finite total energy tell that $\varphi$ cannot be proportional to $\ln(h)$.
By summarizing the above discussion, we obtain,
\begin{align}
\label{homsolconst}
\varphi^{(1,0)}_\mathrm{Hom} = \mathrm{const}\,.
\end{align}
{
We should note that the expression in (\ref{Hom-sol-1}) diverges at the horizon, where $r-M+\sqrt H$ vanishes.
The above arguments about the boundary conditions, however, tell that the homogeneous solution
must be a constant as shown in Eq.~(\ref{homsolconst}), and therefore the homogeneous solution does not show the singularity anywhere.
As we will see soon in (\ref{theta-sol-SR111}) and (\ref{phi111}), the expressions of the particular part of the solutions
do not include the factor $r-M+\sqrt H$ even in the integrands and therefore the scalar field $\varphi$ is also regular at the horizon.
}

Although we have the homogenous solution of Eq.~\eqref{1st-eq1}, we need a particular solution $\varphi^{(1,0)}_\mathrm{Part}$
to find the full inhomogeneous solution.
The particular solution is given by,
\begin{align}
\label{theta-sol-SR111}
\varphi^{(1,0)}_\mathrm{Part} \left(r,\theta\right)=&\, \frac{48\epsilon\sigma{ a_\mathrm{KN}}}{\sigma_1 H } \left\{
\sqrt H \left( r-M \right) \int \frac{\left( Mr- q^2 \right) \left( 3Mr-2 q^2 \right)\left[ \left( r-M \right) \tanh^{-1}
\left( {\frac{M-r}{\sqrt H}} \right) +\sqrt H\right] }{r^7}dr \right. \nonumber\\
&\, \left. + \left( \sqrt H \left( r-M \right) \tanh^{-1} \left( {\frac{M-r}{\sqrt H}} \right)
 -H \right)\int \frac{ \left( 3 r^2 M^2-5 q^2Mr+2 q^4 \right) \left( M-r \right) }{r^7}{dr} \right\} \,.
\end{align}
Eq.~\eqref{theta-sol-SR111} which describes the particular solution gives the following form when the charge $q$ vanishes,
\begin{align}
\label{asym}
\left. \varphi^{(1,0)}_\mathrm{Part}\left(r,\theta\right) \right|_{q=0}=\frac{144\epsilon \sigma{ a_\mathrm{KN}}}{M \sigma_1}
&\, \left( \left[M-r\right] \left[\int \frac{\tanh^{-1} \left( \frac{M- r}{M} \right) \left[M- r\right] -M}{r^5} dr \right. \right. \nonumber \\
&\, \left. \left. -\left\{\tanh^{-1} \left( \frac{M-r}{M} \right) - \frac{M}{M-r}\right\}\int \frac{M-r}{ r^5}{dr}\right]\right) \,.\nonumber\\
\end{align}
The asymptotic form of Eq.~\eqref{asym} gives Eq.~(35) in \cite{Yunes:2007ss} , which is derived for slowly uncharged rotating solution
in dCS theory.

Because we have succeeded to present the solution of the dCS scalar field, we try to derive dCS corrections for the metric perturbations.
We stress that the stress-energy tensor~\eqref{Tab-theta} of the dCS scalar field appears
in the modified field equations \eqref{eom} up to $\mathcal{O}(2,1)$,
and thus we neglect the contributions in the metric perturbation.
In such a case, the modified Einstein equations are divided into two types:
The first one constitutes a closed system of the differential equations involving ${h_1}^{(1,1)}$, ${h_2}^{(1,1)}$, ${h_3}^{(1,1)}$, and ${h_4}^{(1,1)}$,
which comes from the components $(t,t)$, $(r,r)$, $\left(r,\theta\right)$, $(\theta,\theta)$ and $(\phi,\phi)$-components.
The second set which comes from the modified Einstein equations yields one differential equation for $\omega^{(1,1)}$,
which is the $(t,\phi)$-component of the modified Einstein equations.

The first type does not depend on the dCS field, $\varphi$, and therefore they are not changed from the equations in GR.
Therefore, we only consider the second set, $(t,\phi)$-component, which gives,
\begin{align}
\label{V1}
&\, \frac{576 \kappa^2 \sigma^2 a_\mathrm{KN}}{r^7 H^3\sigma_1} \left\{H_1 h r^2 \int \frac{ \left( Mr- q^2 \right) \left(3 Mr-2 q^2 \right)
\left[ \sqrt H-\left( M-r \right) \tanh^{-1} \left( \frac{M-r}{\sqrt H} \right)\right]}{r^7}dr+ \left\{ H_1 h r^2 \tanh^{-1} \left( \frac{M-r}{\sqrt H} \right)
\right. \right. \nonumber\\
&\, \left. \left. - \left( \left[ q^2+2 M^2 \right] r^2 -6 M q^2 r + 3 q^4 \right) \sqrt{M^2- q^2} \right\}
\int \frac{ \left( Mr- q^2 \right) \left( 3Mr-2 q^2 \right) \left( M-r \right) }{r^7}dr \right\} \nonumber\\
=&\, r^2 h \omega_{rr} \left( r,\theta \right)+ \omega_{\theta \theta} \left( r,\theta \right) +4r h{\omega}_r \left( r, \theta \right)
+3 \omega_\theta \left( r,\theta \right) \cot \theta \,,
\end{align}
where $H_1=\left( q^2+2 M^2 \right) r-3M q^2$.
When $q=0$, Eq.~\eqref{V1} gives the following form,
\begin{align}
\label{111}
\omega^{(1,1)}_{,\theta\theta} + 3\cot \theta \omega^{(1,1)}_{,\theta} + 4 r h \omega^{(1,1)}_{,r} + r^2h\omega^{(1,1)}_{,rr}
\approx \frac{15 \kappa^2 \mu^2 a_\mathrm{KN} h}{\nu r^8} \sin^2 \theta \left(3 r^2 + 8 M r + 18 M^2\right)\,,
\end{align}
which coincides with the form derived in \cite{Yunes:2007ss} in the case of the uncharged BH.
Again, the general solution is a linear combination of a homogeneous solution and a particular solution.
The particular solution is given by,
\begin{align}
\label{w-sol-SR22}
\omega^{(1,1)} =&\, -\frac{576 \kappa^2 a_\mathrm{KN} \sigma^2}{r^5 H^{3/2} h\sigma_1}
\int \frac{1}{r^4}\left[ \int \left\{ r^2 h H_1 \int \frac{\left( Mr- q^2 \right) \left[ \left(M-r \right) \tanh^{-1} \left( \frac{ M-r}{\sqrt H} \right) -\sqrt H \right]
\left( 3Mr-2 q^2 \right)}{r^7}dr \right. \right. \nonumber\\
&\, -r^2hH_1\left[\tanh^{-1} \left( \frac{M-r}{\sqrt H} \right) - \frac{\left( \left( q^2+2 M^2 \right) r^2-3 q^2 \left[ Mr- q^2 \right] \right)
\sqrt{ H }}{r^2h H} \right] \nonumber \\
&\, \left. \left. \quad \times \int \frac{ \left( Mr- q^2 \right) \left( M-r \right) \left( 3Mr-2 q^2 \right) }{r^7}{dr} \right\} dr \right] dr \,.
\end{align}
The asymptotic form of Eq.~\eqref{w-sol-SR22} has the following form,
\begin{align}
\label{w-sol-SR11}
\omega^{(1,1)} \approx-\frac{ 9 \kappa^2 \sigma^2 a_\mathrm{KN} M \left( 42 M^4+175 M^2 q^2+35 q^4 -80 M^3r -40 q^2 M r \right) }{35 \sigma_1 H^{3/2} r^8}
= \frac{9 \kappa^2 \sigma^2 a_\mathrm{KN}\,M}{35 \sigma_1 H^{3/2} r^7}\left(H_2-\frac{H_3} r \right) \,,
\end{align}
where $H_2=40M \left( 2M^2-q^2 \right)$ and $H_3=42 M^4+175 M^2 q^2+35 q^4$.
Eq.~\eqref{w-sol-SR11} is different from the one derived in \cite{Yunes:2007ss} when $q=0$.
The reason why Eq.~\eqref{w-sol-SR11} is different from the one derived in \cite{Yunes:2007ss} when $q=0$ comes from
the terms including $\tanh^{-1} \left( {\frac{ M-r}{\sqrt H}} \right)$.
The homogeneous solution of Eq.~\eqref{w-sol-SR22} is a sum of generalized hypergeometric functions, whose argument is $r/(2 M)$ and has some separation constant $c_6$.
Although certain values of such constant make the solution purely real, the solution diverges at the spatial infinity.
The other values of constant $c_6$ make the solution infinite or complex.
The aforementioned discussion forces us to choose the integration constants, which are the coefficients of the hypergeometric functions, to vanish.
Therefore, Eq.~\eqref{w-sol-SR22} gives the full solution.

The full gravitomagnetic metric perturbation in the linear order with respect to $\varepsilon$ and $\xi$ yields,
\begin{align}
\label{theta-sol-SR11}
\omega =&\, - \frac{2 M ra_\mathrm{KN}-q^2}{r^4} \nonumber\\
&\, +\frac{576 a_\mathrm{KN} \kappa^2 \sigma^2}{r^5 H^{3/2} h \sigma_1}\int \frac{1}{r^4} \left[ \int
\left\{ r^2 h H_1 \int \frac{\left( Mr- q^2 \right) \left[ \left(M-r \right) \tanh^{-1} \left( \frac{ M-r}{\sqrt H} \right) -\sqrt H \right]
\left( 3Mr-2 q^2 \right)}{r^7} dr \right. \right. \nonumber\\
&\, -r^2hH_1\left[\tanh^{-1} \left( {\frac{M-r}{ \sqrt H}} \right) - \frac{\left( \left( q^2+2 M^2 \right) r^2-3 q^2 \left[Mr- q^2 \right] \right)
\sqrt{ H }}{r^2h H} \right] \nonumber \\
&\, \quad \left. \left. \times \int \frac{ \left( Mr- q^2 \right) \left( M-r \right) \left( 3Mr-2 q^2 \right) }{r^7}dr \right\} {dr}\right]{dr}\,,
\end{align}
whose asymptotic form is given by,
\begin{align}\label{22}
\omega \approx -\frac{2 M ra_\mathrm{KN}-q^2}{r^4} + \frac{9 \kappa^2 \sigma^2 a_\mathrm{KN} M}{35 \sigma_1 H^{3/2} r^7}\left(H_2-\frac{H_3} r \right)\,.
\end{align}
Eq.~\eqref{theta-sol-SR11} constitutes the first charged slow-rotating BH solution in dCS modified gravity.
Note that the perturbation is highly suppressed in the far field limit and decreases as $r^{-7}$, which suggests
that its significance can only be observed in the strong field regime.
Now let us discuss the solution~\eqref{theta-sol-SR11} with/without the charge $q$.
When $q=0$ the solution decreases as $r^{-6}$ \cite{Yunes:2007ss} and when $q\neq 0$ the solution decreases asymptotically as $r^{-7}$.
This means that the solution with $q$ decreases faster than the one without $q$.

We can verify that the approximated solution given by \eqref{theta-sol-SR11} is self-consistent by calculating the next order correction to $\varphi$.
Such correction consists of $\varphi^{(2,0)}$ and $\varphi^{(1,1)}$, which can be calculated by solving the evolution equation to the next order.
Carrying out such calculations, we find,
\begin{align}
\label{phi111}
\varphi^{(1,1)}=&\, \frac{13824 a_\mathrm{KN} \kappa^2 \alpha^3}{\beta^2 H^4} \left( H H_4\left( r-M \right) \int \frac{\left( Mr- q^2 \right) H_4}{r^6}
\left\{\int \frac{1}{r^5 h} \left[ H_1 r^2 h \sqrt H\int \frac{\left( q^2-Mr \right) H_4 \left( 3 Mr - 2 q^2 \right)}{r^7} dr \right. \right. \right. \nonumber\\
&\, -\int \frac{ \left( M-r \right) \left( Mr- q^2 \right) \left( 3Mr-2 q^2 \right) }{r^7} dr \nonumber \\
&\, \quad \times \left. \left. \left( H_1 r^2 h \sqrt H \tanh^{-1} \left( \frac{M-r }{\sqrt H} \right)
+H \left( \left[2 M^2+ q^2 \right] r^2 - 6M r q^2 + 3 q^4 \right) \right) \right] dr \right\} dr \nonumber\\
&\, +\int \frac{\left(M -r \right) \left( Mr- q^2 \right)}{r^6} \left\{ \int \frac{1}{r^5 h} \left[ H_1 r^2h\sqrt H
\int \frac{ \left(q^2- M r \right) H_5 \left( 3 Mr-2 q^2 \right)} {r^7}{dr} \right. \right. \nonumber \\
&\, \quad +\int \frac{ \left( r-M \right) \left( Mr- q^2 \right) \left( 3Mr-2 q^2 \right) }{r^7} dr \nonumber\\
&\, \quad \left. \left. \left. \times\left( H_1 r^2h\sqrt H\tanh^{-1} \left( \frac{M-r}{\sqrt H} \right) +H \left( \left[ 2 M^2+ q^2 \right] r^2
 -6 M r q^2+3 q^4 \right) \right) \right] dr \right\} dr \right)\,,
\end{align}
where $H_4=\left[ \left( r-M \right) \tanh^{-1} \left( {\frac{M-r}{\sqrt H}} \right) +\sqrt H \right]$ and
$H_5=\left[ \left( r-M \right) \tanh^{-1} \left( { \frac{M-r}{\sqrt H}} \right) -\sqrt H \right]$.
Eq.~\eqref{phi111} yields the following asymptotic form,
\begin{align}
\label{theta_111}
\varphi^{(1,1)}\approx&\, - \frac{12 \sigma \xi a_\mathrm{KN} M \cos\theta}{7 \sigma_1 H^3 r^9} \left[\tanh^{-1}\left( \frac{M-r}{\sqrt H} \right)
\left[ r-M \right] - \sqrt H \right] \nonumber \\
&\, \quad \times \left( 2M^2+q^2 \right) \left(63r \left[M^2+q^2 \right]-8M \left[7q^2+9r^2\right] \right) \,.
\end{align}
Eq.~\eqref{theta_111} yields the general behavior of $\varphi^{(1,1)}$ presented in \cite{Yunes:2007ss} when $q=0$.
Eq.~\eqref{theta_111} is $\xi$-times smaller than $\varphi^{(1,0)}$, therefore presenting the small-coupling approximation self-consistent.
To use this improved $\varphi$ solution in the modified field equation, we then find a correction to the metric proportional of order $\xi^2 \epsilon$,
which we neglect in this study.
{ It is important to stress that to obtain the Reissner-Nordstr\"om BH in the static limit, it
is enough to have a constant CS scalar.
Indeed, a constant CS scalar is less restrictive and would allow the topological Pontryagin density in the action which, in turn, is necessary
to set (anti-)self-dual configurations as a ground state \cite{Miskovic:2009bm}. }

\section{Properties of the derived BH solution}\label{properties}
In the following subsections, we will discuss some physical properties of the solution derived in the previous section.

\subsection{Line-element}

{ Gathering the above data, the line element of the slowly rotating BH can be written up to $\mathcal{O}\left({a_\mathrm{KN}}^2\right)$ as
\begin{align}
\label{line}
ds^2 =&\, -g_{tt}dt^2+g_{rr}dr^2+g_{\theta\theta}d\theta^2+g_{\phi\phi} d\phi^2-2g_{t\phi}dtd\phi\, , \\
\varphi^{(1,0)}=&\, \frac{48\epsilon\sigma{ a_\mathrm{KN}}}{\sigma_1 H } \left\{
\sqrt H \left( r-M \right) \int \frac{\left( Mr- q^2 \right) \left( 3Mr-2 q^2 \right)\left[ \left( r-M \right) \tanh^{-1}
\left( {\frac{M-r}{\sqrt H}} \right) +\sqrt H\right] }{r^7}dr \right. \nonumber\\
&\, \left. + \left( \sqrt H \left( r-M \right) \tanh^{-1} \left( {\frac{M-r}{\sqrt H}} \right)
 -H \right)\int \frac{ \left( 3 r^2 M^2-5 q^2Mr+2 q^4 \right) \left( M-r \right) }{r^7}{dr} \right\} \,,
\end{align}
with
\begin{align}
\label{sol:metric_elements}
g_{tt} =&\, -h - \frac{{a_\mathrm{KN}}^2 \left( 2Mr-q^2 \right)}{r^4} \cos^2{\theta}\, , \nonumber \\
g_{t\phi}=&\, \, - \frac{ \left(2 Mr-q^2 \right) a_\mathrm{KN}}{r^2} \sin^2{\theta} \nonumber \\
&\, -\frac{576 a_\mathrm{KN} \kappa^2 \sigma^2}{r^5 H^{3/2} h\sigma_1}
\int \frac{1}{r^4} \left[\int \left\{ r^2 h H_1 \int \frac{\left( Mr- q^2 \right) \left[ \left(M-r \right) \tanh^{-1} \left( \frac{ M-r}{\sqrt H} \right) -\sqrt H\right]
\left( 3Mr-2 q^2 \right)}{r^7} dr \right. \right. \nonumber\\
&\, -r^2 h H_1 \left[\tanh^{-1} \left( {\frac{M-r}{\sqrt H}} \right) - \frac{\left( \left( q^2+2 M^2 \right) r^2-3 q^2 \left[ Mr- q^2 \right] \right)
\sqrt{ H }}{r^2h H} \right] \nonumber \\
&\, \quad \times \left. \left. \int \frac{\left( Mr- q^2 \right) \left( M-r \right) \left( 3Mr-2 q^2 \right) }{r^7}dr \right\} dr \right] dr\,,\nonumber\\
g_{rr} =&\, \frac{1}{h} + \frac{{a_\mathrm{KN}}^2}{h r^2} \left(\cos^2 \theta - \frac{1}{h} \right)\, , \nonumber \\
g_{\theta \theta} =&\, r^2 + {a_\mathrm{KN}}^2 \cos^2 \theta \, , \nonumber \\
g_{\phi \phi} =&\, r^2 \sin^2 \theta + {a_\mathrm{KN}}^2 \sin^2 \theta \left(1 + \frac{2 Mr-q^2}{r^2} \sin^2 \theta \right)\, ,\nonumber \\
A_t =&\, -\frac q r +\frac{q \sin^2\theta a_\mathrm{KN}} r \,.
\end{align}
Eq.~\eqref{sol:metric_elements} is correct in the orders of $\mathcal{O}(2,0)$, $\mathcal{O}(1,1)$, and $\mathcal{O}(0,2)$.
We stress that the terms including $\sigma$ and $\sigma_1$ in $g_{t\phi}$ of Eq.~ \eqref{sol:metric_elements} cannot be removed by any coordinate transformation.
We also note that the perturbed parts of the metric are regular at the horizon, where $r-M+\sqrt H$ vanishes.
Therefore as long as we treat the model perturbatively, the scalar invariants are regular at the horizon.
}

{
When $r$ is large i.e., $r\to \infty$, the Pontryagin density $R_{\nu\mu\alpha\beta} \tilde R^{\mu\nu\alpha\beta}$ behaves as:
\begin{align}
\label{inv1}
R_{\nu\mu\alpha\beta} \tilde R^{\mu\nu\alpha\beta}\approx&\, -\frac{576 \epsilon a_\mathrm{KN} M^2 \cos\theta }{r^7}
+\frac{960 \epsilon a_\mathrm{KN} Mq^2 \cos\theta}{r^8}-\frac{384\epsilon a_\mathrm{KN} q^4 \cos\theta }{r^9} \nonumber \\
&\, +\frac{432\epsilon \sigma^2a_\mathrm{KN} M^3 (2M^2+q^2)\cos\theta }{\sigma_1 (M^2-q^2)^{3/2}r^{11}}+\mathcal{O}\left(\frac{1}{r^{12}}\right) \,.
\end{align}
Eq.~\eqref{inv1} shows the corrections coming from the CS term to the order under consideration in the present study.
The Pontryagin density $ R \tilde R $, which is proportional to $\square \varphi$, and its deviation from that
in the Kerr-Newman can be calculated by using \eqref{phi111}.
Also, Eq.~\eqref{inv1} shows the correction of the CS scalar field starts in the invariant $R_{\nu\mu\alpha\beta} \tilde R^{\mu\nu\alpha\beta}$
from $\mathcal{O}\left(\frac{1}{r^{11}}\right)$ and the lower orders appear in this invariant are due to the contribution coming from the slowly rotating Kerr-Newman BH.
Also, Eq.~\eqref{inv1} tells that the cross term of the line element \eqref{line} cannot vanish or be gauged away by any coordinate transformation.
This is because if the cross term in \eqref{line} vanishes, $a_\mathrm{KN}=0$ and in that case, the CS scalar field given
by Eq.~\eqref{asym} also vanishes and the CS theory reduces to GR of the Einstein theory. }

The metric in Eq.~\eqref{sol:metric_elements} involves a true singularity at $r=0$.
This can be investigated by calculating the Kretchmann invariant $R_{\mu\nu\rho\sigma} R^{\mu\nu\rho\sigma}$, the squared of Ricci tensor $R_{\mu\nu} R^{\mu\nu}$,
and the Ricci scalar $R$, which ensures the divergence at $r=0$.
Moreover, Eq.~\eqref{inv1} shows also that the Pontryagin density and the CS scalar field diverge at $r=0$.

The location of the event horizon can be found by solving the equation $g_{tt} g_{\phi\phi} - g_{t\phi}^2 =0$, which yields,
\begin{align}
r_\mathrm{H,KN}=r_\mathrm{KN} =M+\sqrt{M^2 - {a_\mathrm{KN}}^2-q^2} \,.
\label{hor}
\end{align}

Moreover, the location of the ergosphere can also be derived by solving the equation $g_{tt}=0$ for $r$,
\begin{align}
\label{ergo}
r_\mathrm{ergo} = r_\mathrm{ergo,KN}\,,
\end{align}
with the ergosphere of the Kerr-Newmann solution given by $r_\mathrm{ergo, KN} = M + \sqrt{M^2 - {a_\mathrm{KN}}^2 \cos^2 \theta-q^2}$.
Eq.~(\ref{ergo}) tells that the radius of the ergosphere is not changed from that in the Kerr-Newmann solution.

The choice of the homogeneous integration constants depends on how to choose the definition of the mass $M$ and the reduced spin angular momentum $a_\mathrm{KN}$.
Therefore, a natural choice is to define such quantities so that they coincide with those measured by an observer at $r\to \infty$ and we obtain the metric displayed
in (\ref{sol:metric_elements}).
With these definitions, the angular velocity and area of the event horizon are changed from those in the Kerr-Newmann solution as follows,
The new solution given by Eq.~\eqref{sol:metric_elements} amended the dragging of the inertial frame of the rotation of the BH.
This can be calculated by the angular velocity $\omega_Z$ for the zero-angular-momentum observer, which is defined by,
\begin{align}
\omega_Z = -\frac{g_{t\phi}}{g_{\phi\phi}}\,,
\end{align}
which yields,
\begin{align}
\label{ang-mom1}
\omega_Z =&\, - \frac{2 M a_\mathrm{KN}}{r^3}\nonumber\\
&\, +\frac{576 \kappa^2 a_\mathrm{KN} \sigma^2}{r^5 H^{3/2}h \sigma_1}
\int \frac{1}{r^4}\left[ \int \left\{ r^2 h H_1 \int \frac{\left( Mr- q^2 \right) \left[ \left(M-r \right) \tanh^{-1} \left( {\frac{ M-r}{\sqrt H}} \right)
 -\sqrt H \right] \left( 3Mr-2 q^2 \right)}{r^7}{dr} \right. \right. \nonumber\\
&\, -r^2hH_1\left[\tanh^{-1} \left( \frac{M-r}{ \sqrt H} \right)
 - \frac{\left( \left( q^2+2 M^2 \right) r^2-3 q^2 \left[ Mr- q^2 \right] \right) \sqrt{ H }}{r^2h H} \right] \nonumber \\
&\quad \times \left. \left. \int { \frac{\left( Mr- q^2 \right) \left( M-r \right) \left( 3Mr-2 q^2 \right) }{r^7}} dr \right\} dr \right] dr\,.
\end{align}
{ Now we are in a position to discuss the conserved charges of the above solution.
We use the following transformation to transform the metric (\ref{sol:metric_elements})  to
the Cartesian coordinates via the following standard transformation \cite{Yunes:2009hc}:
\begin{align}
\label{trans}
x =&\, r \left(1 + \frac{a_\mathrm{KN}^{2}}{2 r^{2}} \right) \cos{\phi} \sin{\theta}\, ,
\nonumber \\
y =&\, r \left(1 + \frac{a_\mathrm{KN}^{2}}{2 r^{2}} \right) \sin{\phi} \sin{\theta}\, ,
\nonumber \\
z =&\, r \cos{\theta}\, ,
\end{align}
We also define $\bar h_{\mu \nu}$ by the difference between ${g}_{\mu \nu}$ in (\ref{sol:metric_elements}) and the flat metric $\eta_{\mu \nu}$,
\begin{align}
\label{back}
\bar h_{\mu \nu}\equiv g_{\mu \nu} - \eta_{\mu \nu} \, .
\end{align}
We should note that $\bar h_{\mu \nu}$ needs not to be small everywhere but we require that
at infinity, away from the black hole, it goes to vanish because we are considering the asymptotically flat spacetime as a solution.
A detailed discussion on how one can use Eq.~ (\ref{back}) to derive the mass formula for the CS-modified GR in the asymptotically
flat spacetime can be found in \citep{Tekin:2007rn}.
Here we list the mass formula of Chern-Simons-modified GR as in \cite{Tekin:2007rn}:
\begin{align}
E = \frac{1}{16 \pi G} \oint_{S^2} dS_{i} \, \left( Q^{0 i}_{E} \left(\bar{\xi} \right) + \frac{1}{2}
Q^{0 i}_{E} \left( \bar{\Xi} \right) - \frac{1}{2} Q^{0 i}_{C} \left(\bar{\xi} \right)\right) \, ,
\label{csgrmass}
\end{align}
where
\begin{align}
\label{einchar}
Q^{\mu i}_{E} \left(\bar{\xi} \right) \equiv &\, \sqrt{-\bar{g}} \left(
\bar{\xi}_{\nu} \bar{\nabla}^{\mu} \bar h^{i \nu} - \bar{\xi}_{\nu} \bar{\nabla}^{i} \bar h^{\mu\nu}
+ \bar{\xi}^{\mu} \bar{\nabla}^{i} \, \bar h - \bar{\xi}^{i} \bar{\nabla}^{\mu} \bar h \right. \nonumber \\
&\, \left. + h^{\mu\nu} \bar{\nabla}^{i} \bar{\xi}_{\nu} -\bar h^{i \nu} \bar{\nabla}^{\mu} \bar{\xi}_{\nu}
+ \bar{\xi}^{i} \bar{\nabla}_{\nu} \bar h^{\mu\nu} - \bar{\xi}^{\mu} \bar{\nabla}_{\nu} \, \bar h^{i \nu}
+ \bar h \, \bar{\nabla}^{\mu} \, \bar{\xi}^{i} \right) \, , \nonumber \\
Q^{\mu i}_{C} \left(\bar{\xi} \right) \equiv &\, v_\sigma \bar{\xi}^\rho \epsilon^{\sigma \mu i \beta} {\mathcal{G}^L}_{\rho \beta}
+ v_\sigma \bar{\xi}_\nu \epsilon^{\sigma \nu i \beta} {\mathcal{G}^{L\mu}}_\beta
+ v_\sigma \bar{\xi}_\nu \epsilon^{\sigma \mu \nu \beta} {\mathcal{G}^{Li}}_\beta\,,
\end{align}
and the integration is performed on a two-dimensional spacial sphere $S^2$ with a large enough radius.
In Eq.~(\ref{einchar}), $\bar h$ is the trace of $\bar h^{\mu\nu}$, i.e., $\bar h=g_{\mu\nu}\bar h^{\mu\nu}$,
${\mathcal{G}^L}_{\mu\nu}$ is the linearized form of the Einstein tensor,
\begin{align}
\label{LEtensor}
{\mathcal{G}^L}_{\mu\nu} = \frac{1}{2} \left( - \bar\Box {\bar h}_{\mu\nu} - \bar\nabla_\mu \bar\nabla_\nu \bar h
+ \bar\nabla^\sigma \bar\nabla_\nu {\bar h}_{\sigma\mu}
+ \bar\nabla^\sigma \bar\nabla_\mu {\bar h}_{\sigma\nu}\right)
 - \frac{1}{2} {\bar g}_{\mu\nu} \left( \bar\Box \bar h + \bar\nabla^\rho \bar\nabla^\sigma {\bar h}_{\rho\sigma} \right)  \, ,
\end{align}
and $\bar{\nabla}^{\mu}$ is the covariant derivative with respect to the background.
Because we are considering the flat background (\ref{back}), we find that the covariant derivatives reduce to the partial derivatives,
$\bar{\nabla}^{\mu}=\partial^\mu$.
In Eq.~(\ref{einchar}), $v_\sigma$ is defined in Eq.~(\ref{v}).
Now let us apply the formula of energy given by Eq.~(\ref{csgrmass}) to the BH~(\ref{sol:metric_elements}).
For the energy, $\bar{\xi}^\mu$ is the time-like Killing vector defined as $\bar{\xi}^\mu = ( -1, 0, 0, 0)$.
An explicit computation of the mass of this metric using the formula~(\ref{csgrmass}) is straightforward.
$\bar{\Xi}^\mu$ vanishes\footnote{ Here $\bar{\Xi}^\mu$ is defined as
\[
\bar{\Xi}^\mu \equiv \frac{1}{\sqrt{-g}} v_\sigma \epsilon^{\sigma \nu \alpha \mu}
\bar{\nabla}_\alpha \bar{\xi}_\nu\,.
\]}
so $Q_E \left( \bar{\Xi} \right)$ term does not contribute.
Moreover, $Q_C\left(\bar{\xi} \right)$ term vanishes for various reasons (such as symmetry and because ${\mathcal{G}^{L\mu}}_\beta$ vanishes for this Einstein space at infinity).
 From the first part, we obtain $E = M$ at infinity up to order $a_\mathrm{KN}$.
Therefore there are no corrections from the CS term.

Now we are ready to write down the formula that enables us to calculate the angular momentum of the Chern-Simons-modified GR for asymptotically
flat space and generalizes Eq.~(\ref{csgrmass}).
This formula takes the following form (for more details, see \citep{Tekin:2007rn}),
\begin{equation}
\label{ang}
Q^0({\xi}_\mu) = \frac{1}{16 \pi G} \oint_{S^2} dS_i \left[ \bar{\xi}_0 (\partial_j \bar h^{ij} - \partial^i \bar h^j\,_j) + \bar{\xi}^i\partial_j \bar h^{0j}
 - \bar{\xi}_j\partial^i \bar h^{0 j} +\frac{\sigma_1}{2}\epsilon^{i j k}\bar{\xi}_j{\mathcal{G}^{L0}}_k \right]\, .
\end{equation}
Eq.~(\ref{ang}) for $\bar{\xi}_0 = ( 1, 0, 0,0)$ gives  $Q^0({\xi}_0)= E$ and the formula (\ref{ang}) coincides with the usual ADM one.
As Eq.~ (\ref{ang}) shows, the effect of the CS term, which appears in the last term, does not contribute.


In the case of the angular momentum, we consider the case of $\bar{\xi}_i = ( 0, 0, 0, 1 )$ and $Q^0 \left( \bar{\xi}_i \right)= J$.
The first three terms in the integrand of (\ref{ang}) are identical to those in the Einstein gravity.
Because Eq.~(\ref{sol:metric_elements}) tells that the corrections coming
from the CS term decrease rapidly compared with the terms appearing in the Kerr geometry in the Einstein gravity, they do not contribute
to the angular momentum.
For the last term $\epsilon^{i j k}\bar{\xi}_j{\mathcal{G}^{L0}}_k$, which appears due to the existence of the CS term,
the $i$-direction is perpendicular to the two-dimensional surface $S^2$ and therefore
the $i$-direction corresponds to the radial ($r$) direction.
On the other hand, ${\bar \xi}_j$ is a unit vector corresponding to the $z$-direction.
This tells that the $k$-direction corresponds to the $\phi$-direction and therefore only ${\mathcal{G}^L}_{t\phi}$ contributes to $Q^0(\bar{\xi}_i)$.
Because Eq.~(\ref{sol:metric_elements}) tells that the metric does not depend on the time $t$ nor angular $\phi$,  all the terms
except the first term $- \frac{1}{2} \bar\Box {\bar h}_{\mu\nu}$ in the expression~(\ref{LEtensor}) corresponding to ${\mathcal{G}^L}_{t\phi}$ vanish trivially.
Eq.~(\ref{sol:metric_elements}) also tells that ${\bar h}_{t\phi}$ is $\mathcal{O}\left( r^{-1} \right)$ and therefore $\bar\Box {\bar h}_{t\phi} \sim \mathcal{O}\left(r^{-3}\right)$.
Because the area of the two-dimensional surface $S^2$ is $\mathcal{O}\left( r^2 \right)$, the last term in (\ref{ang}) for $Q^0(\bar{\xi}_i)= J$ does not contribute to $J$
in the limit that the radius of $S^2$ goes to infinity.
Therefore there is no correction from the CS term to the angular momentum and we obtain,
\begin{align}
\label{angc}
J=a_\mathrm{KN} M \, ,
\end{align}
which is identical to the angular momentum in the standard Kerr(-Newman) black hole.
Eq.~(\ref{angc}) shows that the effect of the CS scalar field on the calculation of the angular momentum vanishes up to $a_\mathrm{KN}$.}

{ Now we are going to discuss the Hawking temperature of the BH solution given by Eq.~(\ref{sol:metric_elements}).
The Hawking temperature $T$ is generally defined by the surface gravity $\kappa$ so that $T=\kappa/\left(2\pi\right)$ and we now obtain
\cite{Sheykhi:2012zz,Sheykhi:2010zz,Hendi:2010gq,Sheykhi:2009pf},
\begin{align}
\label{temp11}
T_\mathrm{H,KN} = \frac{h' \left( r_\mathrm{H,KN} \right)}{4\pi}\,.
\end{align}
Using Eq.~(\ref{sol:metric_elements}) in Eq.~(\ref{temp11}), we obtain the Hawking temperature as,
\begin{align}
\label{temp111}
T_\mathrm{H,KN}=\frac{{r_\mathrm{H,KN}}^2-q^2}{4\pi {r_\mathrm{H,KN}}^3}\,,\quad \mbox{up to} \quad \mathcal{O}\left(a_\mathrm{KN}\right)\,.
\end{align}
Eq.~(\ref{temp111}) does not differ from Reissner-Nordstr\"om solution up to $\mathcal{O}\left(a_\mathrm{KN}\right)$. }

\section{Geodesic precession in the slowly-rotating charged black hole in the dynamical Chern-Simons modified
gravity}\label{geod}
{

In \cite{Harko:2009kj, Sopuerta:2009iy}, the time-like geodesics of the slowly rotating black hole in dynamical Chern-Simons modified gravity were considered.
Sopuerta \textit{et al.}~\cite{Sopuerta:2009iy} investigated the time-like geodesic equations for the massive particles and discovered that in the Chern-Simons modified gravity,
the location of the innermost stable circular orbit (ISCO) and the three physical fundamental frequencies associated with the particle's time $\tau$ are modified.
However, the geodesic precession of orbits around Chern-Simons black holes is only shown numerically for
a few examples in Ref.~\cite{Sopuerta:2009iy}, with no analytic expression for this physical quantity.
Now let us start using the condition $\theta=\pi/2$, which puts the orbits on the equatorial plane.
In such cases, time-like geodesics can be found to take the form,
\begin{align}
u^{t}=&\, \frac{dt}{d\tau}=\frac{Eg_{\phi\phi}-Lg_{t\phi}}{g^2_{t\phi}+g_{tt}g_{\phi\phi}}\, ,
\label{u1}\\
\mathrm u^{\phi}=&\, \frac{d\phi}{d\tau}=\frac{Eg_{t\phi}+Lg_{tt}}{g^2_{t\phi}+g_{tt}g_{\phi\phi}}\, ,
\label{u2}\\
&\left( \frac{dr}{d\tau}\right)^2+ V_\mathrm{eff}(r)=E^2\, ,
\end{align}
with the effective potential
\begin{align}
V_\mathrm{eff}(r)=\frac{1}{g_{rr}}\left(1+\frac{E^2[g_{rr}(g^2_{t\phi}+g_{tt}g_{\phi\phi})-g_{\phi\phi}]+2ELg_{t\phi}+L^2g_{tt}}{g^2_{t\phi}+g_{tt}g_{\phi\phi}}\right)\, ,
\end{align}
with $E$ and $L$ being the specific energy and angular momentum of particles moving in the orbits, respectively.
The effective potential $V(r)$ must obey for a stable circular orbit in the equatorial plane the following equation,
\begin{align}
V_\mathrm{eff}(r)=E^2\, , \quad \frac{dV_\mathrm{eff}(r)}{dr}=0\, .
\end{align}
By solving the above equations, one obtains,
\begin{align}
E=&\, \frac{g_{tt}+g_{t\phi}\Omega}{\sqrt{g_{tt}+2g_{t\phi}\Omega-g_{\phi\phi}\Omega^2}}
=-{\frac{27a_\mathrm{KN}}{5{\xi}\kappa{\beta}r^3 \left( r^2-2\,Mr+ q^2- r^4\Omega ^2 \right)^{3/2} \left({M}^2 - q^2 \right)^{3/2}}} \nonumber\\
&\, \times\left( {\frac {5}{27}} \kappa \xi \left(q^2- M^2 \right) \beta r^2 \left( 2M\Omega ^3
\epsilon r^{5}-\Omega ^3\epsilon\, q^2 r^4 \right)\sqrt {{M}^2- q^2} \right. \nonumber \\
&\, \left. +\epsilon M\Omega ^3{\alpha}^2
\left[ r \left( {\frac {40}{21}}\,{M}^3+{\frac {20}{21}}M q^2 \right) +{M}^4+\frac{5}{6} q^4+{\frac {25}{6}}{M}^2 q^2 \right]
\right)\nonumber\\
&\, + \frac { \left[ \left(1-\Omega ^2 q^2 \right) r^4- r^{6}\Omega ^2+2M\Omega ^2 r^{5}+ 2\left( 2{M}^2+ q^2 \right) r^2-4 q^2rM
+ q^4-4M r^3 \right] }{\left( r^2-2Mr+ q^2- r^4\Omega ^2 \right)^{3/2} r} \,,\nonumber\\
L=&\, \frac{-g_{t\phi}+g_{\phi\phi}\Omega}{\sqrt{g_{tt}+2g_{t\phi}\Omega-g_{\phi\phi}\Omega^2}}
={\frac {54}{5 r^{7} \left( r^2-Mr+ q^2- r^4\Omega ^2 \right)^{3/2} \left( {M}^2- q^2 \right)^{3/2}{\beta}\kappa{\xi}}}\nonumber\\
&\, \times\left[ {\frac {10}{27}}\kappa\xi\, \left( q^2-M^2 \right) \beta \left( M\epsilon\,\Omega ^2 r^{5}-\frac{\epsilon\,\Omega ^2 q^2 r^4}{2}
- \frac{M r^3\epsilon} + \left( \frac{q^2}{4}+{M}^2 \right) \epsilon r^2-Mr\epsilon q^2
+\frac{\epsilon q^4}{4} \right) r^{6}\sqrt {{M}^2- q^2} \right.\nonumber\\
&\, \left. + M{\alpha}^2\epsilon \left( Mr+ r^4\Omega ^2-\frac{ q^2}{2}-\frac{ r^2}{2} \right)\left[ \left( {\frac {40}{21}}{M}^3
+{\frac{20}{21}}M q^2 \right) r+{M}^4+\frac{5 q^4}{6}+{\frac {25}{6}}{M}^2 q^2 \right] \right] a_\mathrm{KN} \nonumber \\
&\, -{\frac{ \left( r^{6}\Omega - r^{8}\Omega ^3-2M\Omega r^{5}+\Omega q^2 r^4 \right) }{ \left( r^2-2Mr+ q^2- r^4\Omega ^2
\right)^{3/2}r}}\,,\nonumber\\
\Omega=&\, \frac{d\phi}{dt}=\frac{g_{t\phi,r}+\sqrt{(g_{t\phi,r})^2+g_{tt,r}g_{\phi\phi,r}}}{g_{\phi\phi,r}}
=\frac{\epsilon a_\mathrm{KN}}{\left( {M}^2- q^2 \right)^{3/2}{\beta}\kappa{\xi} r^{10}} \left[ \left( \left( {M}^2- q^2 \right)^{3/2}
\beta\kappa\,\xi r^{7}-18\,{\alpha}^2 q^4 \right) M \right.\nonumber\\
&\, \left. -{\frac{108}{5}}{M}^{5}{\alpha}^2-36 {M}^4{\alpha}^2r-90{M}^3{\alpha}^2 q^2-18{M}^2{\alpha}^2 q^2r- q^2
\left( - q^2+{M}^2 \right)^{3/2}\beta\kappa \xi r^{6} \right]+\frac{\sqrt {Mr- q^2}}{{r^2}}\,,
\label{jsd}
\end{align}
with $\Omega$ being the angular velocity of the particle moving in the orbits.

Using Eq.~(\ref{jsd}), one obtains Kepler's third law in the slowly-rotating black-hole spacetime in the dynamical
Chern-Simons modified gravity
\begin{align}
T^2=&\, 4\pi^2\left\{\frac { R^{8}}{ \left( \sqrt { q^4+M R^{5}- R^4 q^2}+ q^2 \right)^2}
+\frac {R^4M {a_\mathrm{KN}}}{35 \left( \sqrt { q^4+M r^{5}- R^4 q^2}+ q^2 \right)^3{\sigma1} H^{3/2}
\left(  q^4+M R^{5}- R^4 q^2 \right) } \right. \nonumber\\
&\, \times \left( 70  R^{5}{\sigma_1} H^{3/2} q^4 -54\,H_3\,\sigma^2\kappa^2q^4-54 H_3 \sigma^2\kappa^2M R^5
+54H_3\sigma^2\kappa^2 R^4 q^2+70 R^{10}{\sigma_1} H^{3/2}M \right. \nonumber \\
&\, -70 R^{9}{\sigma_1} H^{3/2} q^2+45 \kappa^2\sigma^2{H_2} R q^4 +45 \kappa^2\sigma^2H_2\, R^{6}M
-45\kappa^2\sigma^2H_2 R^{5} q^2-54 \sqrt { q^4+M R^{5}- R^4 q^2}{q}^2H_3\,\sigma^2\kappa^2 \nonumber \\
&\, \left. \left. +45 R\sqrt { q^4+M R^{5} - R^4 q^2} q^2\kappa^2\sigma^2H_2
+70 R^5 \sqrt { q^4+M R^{5}- R^4 q^2} q^2{\sigma_1}\, H^{3/2} \right)
+\mathcal{O} \left({a_\mathrm{KN}}^2 \right)\right\}\,,
\label{Time}
\end{align}
where $R$ representing the orbital radius and $T=\frac{1}\Omega $ representing the orbital period.
The subsequent terms in the right-hand side of Eq.~(\ref{Time}) is the correction by the $a_\mathrm{KN}$ and the Chern-Simons term.
The correction term disappears as $a$ approaches zero.
This is sensible because as $a_\mathrm{KN}$ approaches zero the metric (\ref{line}) coincides with
one of the Schwarzschild black holes in general relativity.
Since the black hole rotates slowly, the first-order terms in $a_\mathrm{KN}$ dominate the correction.
As a result, when the black hole rotates in the same direction as the particle, i.e., $a_\mathrm{KN}>0$, the orbital
period $T$ decreases with the Chern-Simons coupling parameter $\xi$.
However, when the black hole rotates in the opposite direction as the
particle, i.e., $a_\mathrm{KN}<0$, the orbital period $T$ tends to increase with the Chern-Simons coupling parameter $\xi$.}

\subsection{Stability of the BH given by Eq.~(\ref{sol:metric_elements}) through the use of geodesic deviation}\label{S666}

The trajectory of a test particle in the BH space-time is prescribed by the geodesic equations of the following form,
\begin{align}
\label{ge}
0= \frac{d^2 x^\alpha}{ d\tau^2} + \left\{ \begin{array}{c} \alpha \\ \mu \nu \end{array} \right\}
\frac{d x^\mu}{d\tau} \frac{d x^\nu}{d\tau} \, ,
\end{align}
where $\tau$ is the affine parameter along the geodesic.
The equation of geodesic deviation has the form \cite{dInverno:1992gxs}:
\begin{align}
\label{ged}
0= \frac{d^2 \varepsilon^\alpha}{d\tau^2} + 2\left\{ \begin{array}{c} \alpha \\ \mu \nu \end{array} \right\}
\frac{d x^\mu}{d\tau} \frac{d \varepsilon^\nu}{d\tau} + \left\{ \begin{array}{c} \alpha \\ \mu \nu \end{array} \right\}_{,\ \rho}
\frac{d x^\mu}{d\tau} \frac{d x^\nu}{d\tau}\varepsilon^\rho \, ,
\end{align}
with $\varepsilon^\rho$ being the deviation 4-vector.
Applying (\ref{ge}) and (\ref{ged}) into \eqref{sol:metric_elements} ,
we obtain the geodesic equations in the following form,
\begin{align}
0=&\, \frac{d^2 t}{d\tau^2}+\epsilon\left[Ma_\mathrm{KN}-\frac{a_\mathrm{KN}q^2}{2r}+r^3A_t \omega\right]\left(\frac{d \phi}{d\tau}\right)^2\left\{1-\frac{h'}{2r}\right\}\, , \nonumber \\
0=&\, \frac{1}{2}h'(r)\left( \frac{dt}{d\tau}\right)^2 -r\left(\frac{d\phi}{d\tau}\right)^2=0\, ,\quad
0= \frac{d^2\theta}{d\tau^2} \, , \quad 0= \frac{d^2 \phi}{d\tau^2} \, .
\end{align}
Using the circular orbit
\begin{align}
\theta = \frac{\pi}{2}\, , \quad \frac{d\theta}{d\tau}=0\, , \quad \frac{dr}{d\tau}=0\, ,
\end{align}
we obtain the geodesic deviation in the following form,
\begin{align}
\label{ged22}
0= &\, \frac{d^2 \varepsilon^0}{d\tau^2} +\frac{h'}{h}\frac{dt}{d\tau} \frac{d\varepsilon^1}{d\tau}
+\epsilon \left[ Ma_\mathrm{KN}-\frac{a_\mathrm{KN}q^2}{2r}+r^3A_t \omega \right]
\left( 2\frac{d \phi}{d\tau} \frac{d \varepsilon^3}{d\tau}-\frac{h'} r \frac{dt}{d\tau} \frac{d \varepsilon^0}{d\tau}\right) \nonumber \\
&\, +\left\{\epsilon\left[A_t r^2 \left( 3\omega+r \omega' \right)+\frac{a_\mathrm{KN}q^2}{2}r^2\right] \left( \frac{d\phi}{d\tau}\right)^2 \right.\nonumber \\
&\, \left. -\frac{\epsilon}{2r^2} \left[2 r^3h'\omega+r^4h'\omega'-Ma_\mathrm{KN}h'+h''Ma_\mathrm{KN}r+\xi h''r^4\omega+\frac{a_\mathrm{KN}q^2h'} r
 -\frac{a_\mathrm{KN}q^2h''}{2}\right]\left( \frac{dt}{d\tau}\right)^2 \right\}\varepsilon^1 \, , \nonumber\\
0= &\, 2 \frac{d^2 \varepsilon^1}{d\tau^2}-4hr \frac{d\phi}{d\tau}\frac{d\varepsilon^3}{d\tau}+2hh' \frac{dt}{d\tau}\frac{d\varepsilon^0}{d\tau}
 -\left\{2 \left[ h+rh' \right]\left( \frac{d\phi}{d\tau}\right)^2- \left[ hh''+h'^2 \right]\left( \frac{dt}{d\tau}\right)^2\right\}\varepsilon^1 \,,\nonumber\\
0=&\, \frac{d^2 \varepsilon^2}{d\tau^2} + \left( \frac{d\phi}{d\tau} \right)^2 \varepsilon^2\,, \quad
0=\frac{d^2 \varepsilon^3}{d\tau^2} +\frac{2} r \frac{d\phi}{d\tau} \frac{d\varepsilon^1}{d\tau} \, ,
\end{align}
where the functions $h=1-\frac{2M} r +\frac{q^2}{r^2}$ and $\omega$ is defined by the second term of the $g_{t \phi}$
of Eq.~(\ref{sol:metric_elements}) via (\ref{slow-rot-ds2}).

\begin{figure}
\centering
\subfigure[~The plot of $\omega$ w.r.t the radial coordinate $r$. Here we assume the numerical values of $M$, $\sigma$, $\sigma_1$ $q$, and $\epsilon$ as $0.1$, $0.1$, $0.1$, $q=0.001$, $0.1$.]{\label{fig:1a}\includegraphics[scale=0.27]{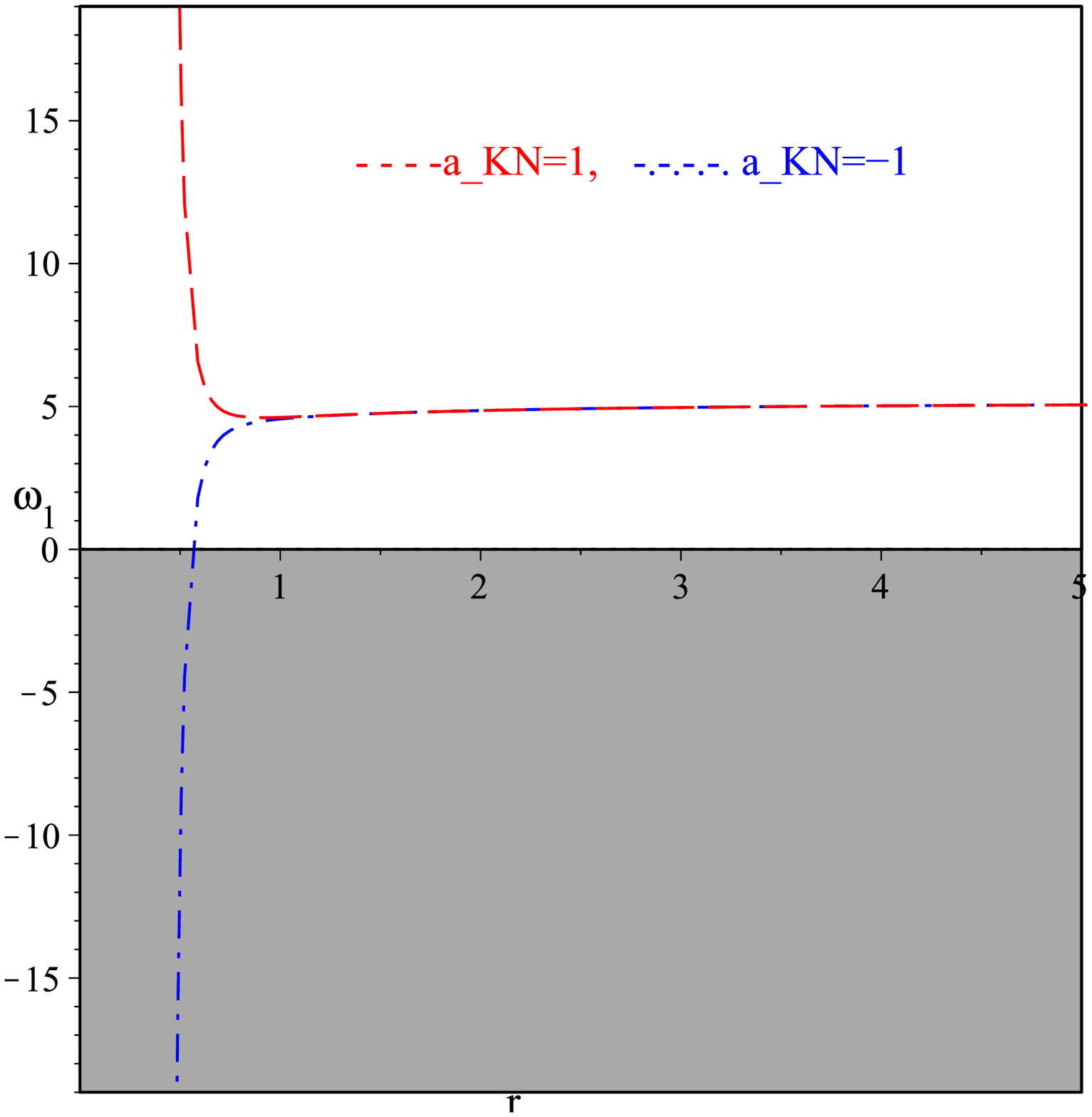}}\hspace{0.2cm}
\subfigure[~The plot of $\omega$ w.r.t the radial coordinate $r$. Here we assume the numerical values of $M$, $\sigma$, $\sigma_1$ $q$, and $a_\mathrm{KN}$ as $0.1$, $0.1$, $0.1$, $q=0.001$, $1$.]{\label{fig:2a}\includegraphics[scale=0.27]{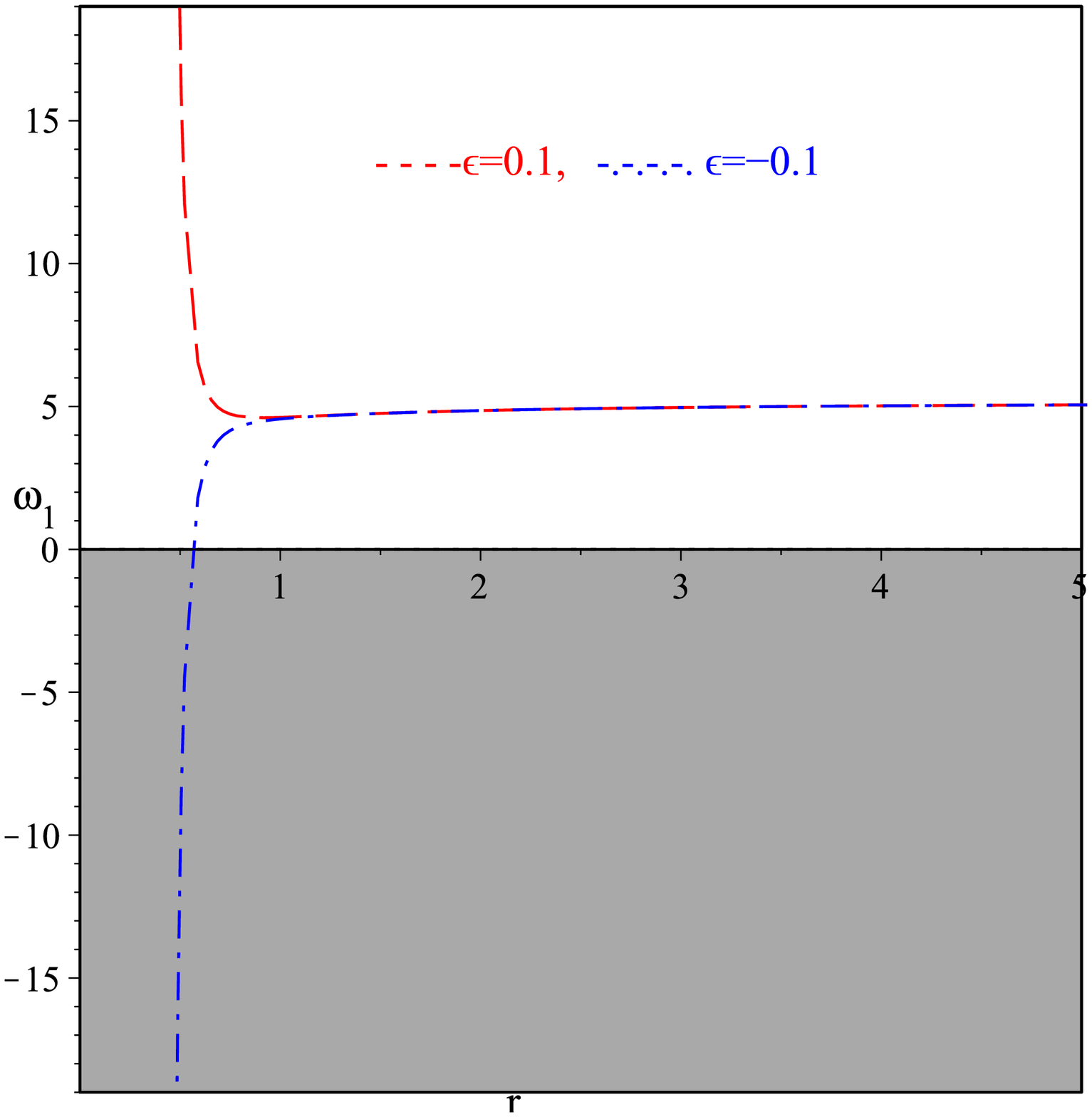}}\hspace{0.2cm}
\subfigure[~The plot of $\omega$ w.r.t the radial coordinate $r$. Here we assume the numerical values of $M$, $\sigma$, $\epsilon$ $q$, and $a_\mathrm{KN}$ as $0.1$, $0.1$, $0.1$, $q=0.001$, $1$.]{\label{fig:3c}\includegraphics[scale=0.27]{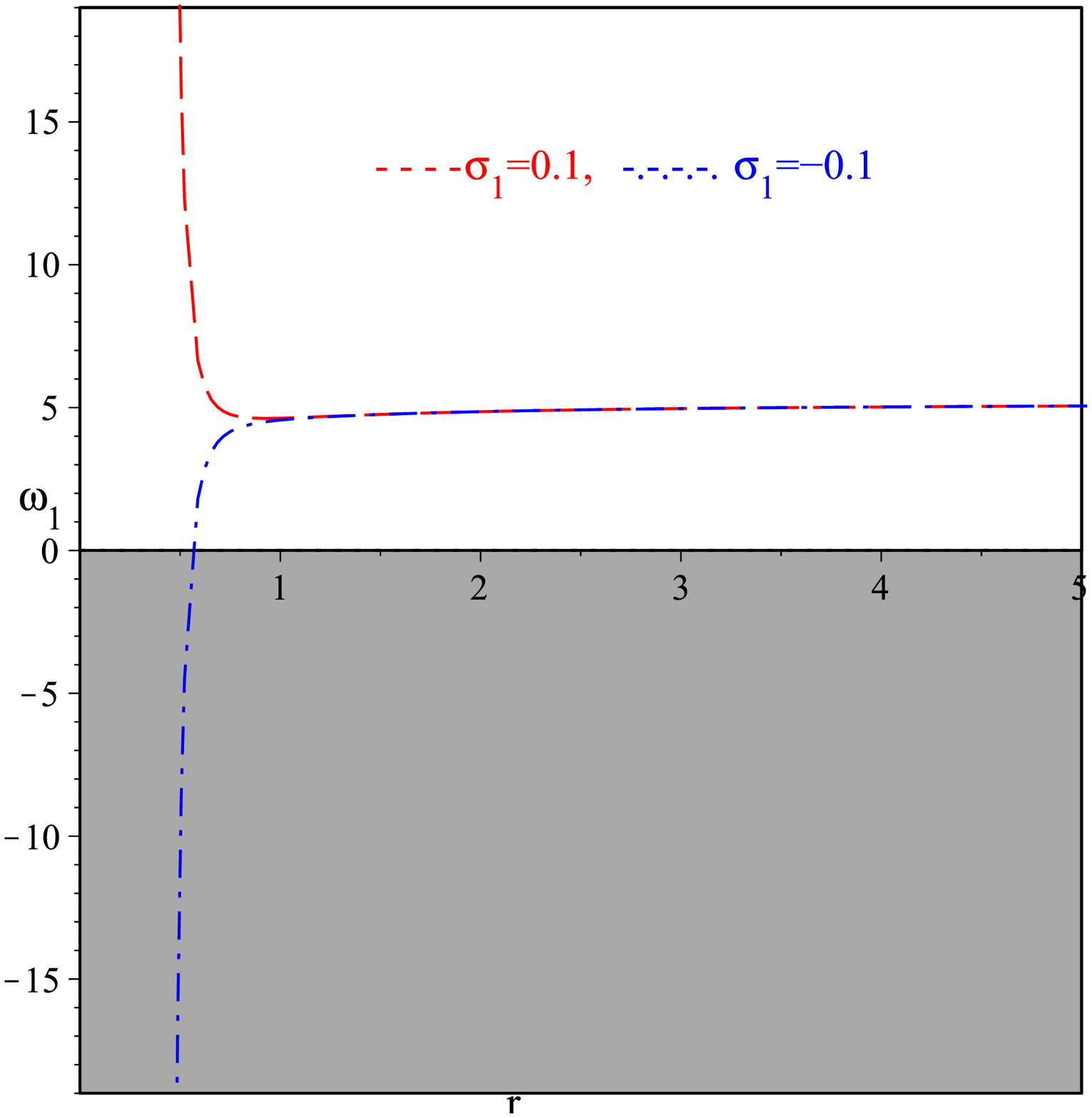}}
\caption{Schematic plots of the $\omega$ with respect to the radial coordinate $r$ of the BH solutions Eqs.~(\ref{sol:metric_elements}) using different values of $a_\mathrm{KN}$, $\epsilon$ and $\sigma_1$.}
\label{Fig:1}
\end{figure}

The third equation of (\ref{ged22}) represents a simple harmonic motion, which ensures that the motion in the plane $\theta=\frac{\pi}{2}$ is stable.
Assuming the solutions of the remaining equations of (\ref{ged22}) to be:
\begin{align}
\label{ged33}
\varepsilon^0 = k_1 \e^{i \omega_1 \phi}\, , \quad \varepsilon^1= k_2\e^{i \omega_1 \phi}\, ,
\quad \mbox{and} \quad \varepsilon^3 = k_3 \e^{i \omega_1 \phi}\, ,
\end{align}
where
$k_1$, $k_2$, and $k_3$ are constants.
Substituting (\ref{ged33}) into (\ref{ged22}), we obtain
\begin{align}
\label{con1}
0=&\, h' \left\{ 2\omega r^3 h' \sqrt{h' r-2 h }-\sqrt{2}\epsilon \left[ \left( \xi r^4\omega +a_\mathrm{KN} \left( Mr- \frac{1}{2} q^2 \right) \right) rh''
 - \left[ \left( r^7\xi-\xi r^5 \right)\omega' -\xi r^4 \left( r^2+2 \right) \omega \right. \right. \right. \nonumber\\
&\, \left. \left. \left. - \frac{a_\mathrm{KN}}{2} \left(2 q^2-5 r^2 q^2-2Mr+8 r^3M \right) \right]h'
 - \left( 2\xi r^5\omega' -2\xi r^4\omega -a_\mathrm{KN} \left(8Mr-5 q^2 \right) \right) h r \right] h \right\} \nonumber \\
&\, - \left( h''h + \left( r^2-1 \right) h'^2 +r \left(5h -{\omega_1}^2 \right)h' -6h^2+2 \omega^2h \right) \nonumber \\
&\, \quad \times \left[ \epsilon h' \sqrt{2} \left( \xi r^4 \omega +a_\mathrm{KN} \left( Mr- \frac{1}{2} q^2 \right) \right)+ r^2\omega_1 \sqrt{ h' r -2 h }\right] r\,.
\end{align}
By solving Eq.~\eqref{con1}, we can derive the form of $\omega_1$, but it is rather tedious.
By using the condition $\omega_1>0$, we can establish the stability condition for the space-times (\ref{sol:metric_elements}) \cite{Misner:1973prb,Nashed:2016tbj,Nashed:2011fg}.
This condition is drawn in Figure~\ref{Fig:1} for different values of the parameters which characterize the space-time (\ref{sol:metric_elements}).

\section{Polarization of Photon near BH}\label{polarization}

We like to consider the possibility to find the BH obtained in this paper by any observation.
In this section, we investigate the propagation of the photon near the BH in (\ref{sol:metric_elements})
and we focus on the difference in polarization from that of the Kerr-Newman BH.

In the general background given by $g_{\mu\nu}=g^{(0)}_{\mu\nu}$ and $A_\mu = A^{(0)}_\mu$,
we may consider the propagation of photon $a_\mu$ which is defined by $a_\mu \equiv A_\mu - A^{(0)}_\mu$
by solving the field equation,
\begin{align}
\label{ph1}
0=\frac{1}{\sqrt{ - g^{(0)}}} \partial_\nu \left\{ \sqrt{ - g^{(0)}} g^{(0)\nu\rho} g^{(0)\mu\sigma}
\left( \partial_\rho a_\sigma - \partial_\sigma a_\rho \right) \right\}\, .
\end{align}
As in (\ref{sol:metric_elements}), we assume
\begin{align}
\label{phi2}
& g^{(0)}_{tt} = - h (r) + \epsilon^2 h_{tt}(r, \theta) \, , \quad
g^{(0)}_{t\phi} = \epsilon h_{t\phi} \left(r,\theta\right) \, , \quad
g^{(0)}_{rr} = \frac{1}{h(r)} + \epsilon^2 h_{rr}(r, \theta) \, , \quad
g^{(0)}_{\theta\theta} = r^2 + \epsilon^2 h_{\theta\theta}(r, \theta) \, , \nonumber \\
& g^{(0)}_{\phi\phi}= r^2 \sin^2 \theta + \epsilon^2 h_{\phi\phi}(r, \theta) \, , \quad
\mbox{other components}=0\, ,
\end{align}
which gives,
\begin{align}
\label{phi3}
& g^{(0)tt} = - \frac{1}{h (r)} + \epsilon^2 \frac{- r^2 \sin^2 \theta h_{tt}(r, \theta) + h_{t\phi} \left(r,\theta\right)^2}
{h (r)^2 r^2 \sin^2 \theta} \, , \nonumber \\
& g^{(0)t\phi} = \epsilon \frac{h_{t\phi} \left(r,\theta\right)}{h (r) r^2 \sin^2 \theta} \, , \quad
g^{(0)rr} = h(r) - \epsilon^2 {h(r)}^2 h_{rr}(r, \theta) \, , \quad
g^{(0)\theta\theta} = \frac{1}{r^2} - \epsilon^2 \frac{h_{\theta\theta}(r, \theta)}{r^4} \, , \nonumber \\
& g^{(0)\phi\phi}= \frac{1}{r^2 \sin^2 \theta} - \epsilon^2
\frac{h (r) h_{\phi\phi}(r, \theta) + h_{t\phi} \left(r,\theta\right)^2}{ h (r) \left(r^2 \sin^2 \theta \right)^2} \, , \quad
\mbox{other components}=0\, .
\end{align}
Here we have used,
\begin{align}
\label{phi4}
g^{(0)}_{tt} g^{(0)}_{\phi\phi} - {g^{(0)}_{t\phi}}^2 =&\,
 - h (r) r^2 \sin^2 \theta + \epsilon^2 \left( r^2 \sin^2 \theta h_{tt}(r, \theta) - h (r) h_{\phi\phi}(r, \theta) - h_{t\phi} \left(r,\theta\right)^2 \right) \nonumber \\
\frac{1}{g^{(0)}_{tt} g^{(0)}_{\phi\phi} - {g^{(0)}_{t\phi}}^2} =&\,
 - \frac{1}{h (r) r^2 \sin^2 \theta}
 - \epsilon^2 \frac{r^2 \sin^2 \theta h_{tt}(r, \theta) - h (r) h_{\phi\phi}(r, \theta) - h_{t\phi} \left(r,\theta\right)^2 }{\left( h (r) r^2 \sin^2 \theta \right)^2} \, .
\end{align}
When we choose $a_t=0$ gauge condition, Eq.~(\ref{ph1}) has the following forms,
\begin{align}
\label{phi5t}
0=&\, g^{(0)t\phi} g^{(0)tt} {\partial_t}^2 a_\phi
+ \frac{1}{\sqrt{ - g^{(0)}}} \partial_r \left\{ \sqrt{ - g^{(0)}} g^{(0)rr} g^{(0)tt}\partial_t a_r \right\}
+ \frac{1}{\sqrt{ - g^{(0)}}} \partial_\theta \left\{ \sqrt{ - g^{(0)}} g^{(0)\theta\theta} g^{(0)tt}\partial_t a_\theta \right\}\nonumber \\
&\, + g^{(0)\phi\phi} g^{(0)tt} \partial_\phi \partial_t a_\phi \, , \\
\label{phi5r}
0=&\, g^{(0)tt} g^{(0)rr} {\partial_t}^2 a_r
+ g^{(0)t\phi} g^{(0)rr} \partial_t \left( \partial_\phi a_r - \partial_r a_\phi \right)
+ g^{(0)\phi t} g^{(0)rr} \partial_\phi \partial_t a_r \nonumber \\
&\, + \frac{1}{\sqrt{ - g^{(0)}}} \partial_\theta \left\{ \sqrt{ - g^{(0)}} g^{(0)\theta\theta} g^{(0)rr}
\left( \partial_\theta a_r - \partial_r a_\theta \right) \right\}
+ g^{(0)\phi\phi} g^{(0)rr} \partial_\phi \left( \partial_\phi a_r - \partial_r a_\phi \right) \, , \\
\label{phi5theta}
0=&\, g^{(0)tt} g^{(0)\theta\theta} {\partial_t}^2 a_\theta
+ g^{(0)t\phi} g^{(0)\theta\theta} \partial_t \left( \partial_\phi a_\theta - \partial_\theta a_\phi \right) \nonumber \\
&\, + g^{(0)\phi t} g^{(0)\theta\theta} \partial_\phi \partial_t a_\theta
+ \frac{1}{\sqrt{ - g^{(0)}}} \partial_r \left\{ \sqrt{ - g^{(0)}} g^{(0)rr} g^{(0)\theta\theta}
\left( \partial_r a_\theta - \partial_\theta a_r \right) \right\}
+ g^{(0)\phi\phi} g^{(0)\theta\theta} \partial_\phi \left( \partial_\phi a_\theta - \partial_\theta a_\phi \right) \, , \\
\label{phi5phi}
0=&\, g^{(0)tt} g^{(0)\phi\phi} {\partial_t}^2 a_\phi
+ g^{(0)\phi t} g^{(0)\phi\phi} \partial_\phi \partial_t a_\phi \nonumber \\
&\, + \frac{1}{\sqrt{ - g^{(0)}}} \partial_r \left\{ \sqrt{ - g^{(0)}} g^{(0)rr} g^{(0)\phi\phi}
\left( \partial_r a_\phi - \partial_\phi a_r \right) \right\}
+ \frac{1}{\sqrt{ - g^{(0)}}} \partial_\theta \left\{ \sqrt{ - g^{(0)}} g^{(0)\theta\theta} g^{(0)\phi\phi}
\left( \partial_\theta a_\phi - \partial_\phi a_\theta \right) \right\} \, .
\end{align}

We consider a small region and the case the wavelength is small enough compared with the region.
Then the background metric $g_{\mu\nu}=g^{(0)}_{\mu\nu}$ and the background vector field $A_\mu = A^{(0)}_\mu$ can be regarded
to be adiabatically constant in the region.
Under the assumptions, Eqs.~(\ref{phi5t}), (\ref{phi5r}), (\ref{phi5theta}), and (\ref{phi5phi}) in the $a_t=0$ gauge condition have the following forms,
\begin{align}
\label{phi18t}
0=&\, g^{(0)t\phi} g^{(0)tt} {\partial_t}^2 a_\phi + g^{(0)rr} g^{(0)tt} \partial_r \partial_t a_r
+ g^{(0)\theta\theta} g^{(0)tt} \partial_\theta \partial_t a_\theta
+ g^{(0)\phi\phi} g^{(0)tt} \partial_\phi \partial_t a_\phi \, , \\
\label{phi18r}
0=&\, g^{(0)tt} g^{(0)rr} {\partial_t}^2 a_r
+ g^{(0)t\phi} g^{(0)rr} \partial_t \left( \partial_\phi a_r - \partial_r a_\phi \right)
+ g^{(0)\phi t} g^{(0)rr} \partial_\phi \partial_t a_r \nonumber \\
&\, + g^{(0)\theta\theta} g^{(0)rr} \partial_\theta \left( \partial_\theta a_r - \partial_r a_\theta \right)
+ g^{(0)\phi\phi} g^{(0)rr} \partial_\phi \left( \partial_\phi a_r - \partial_r a_\phi \right) \, , \\
\label{phi18theta}
0=&\, g^{(0)tt} g^{(0)\theta\theta} {\partial_t}^2 a_\theta
+ g^{(0)t\phi} g^{(0)\theta\theta} \partial_t \left( \partial_\phi a_\theta - \partial_\theta a_\phi \right) \nonumber \\
&\, + g^{(0)\phi t} g^{(0)\theta\theta} \partial_\phi \partial_t a_\theta
+ g^{(0)rr} g^{(0)\theta\theta} \partial_r \left( \partial_r a_\theta - \partial_\theta a_r \right)
+ g^{(0)\phi\phi} g^{(0)\theta\theta} \partial_\phi \left( \partial_\phi a_\theta - \partial_\theta a_\phi \right) \, , \\
\label{phi18phi}
0=&\, g^{(0)tt} g^{(0)\phi\phi} {\partial_t}^2 a_\phi
+ g^{(0)\phi t} g^{(0)\phi\phi} \partial_\phi \partial_t a_\phi \nonumber \\
&\, + g^{(0)rr} g^{(0)\phi\phi} \partial_r \left( \partial_r a_\phi - \partial_\phi a_r \right)
+ g^{(0)\theta\theta} g^{(0)\phi\phi} \partial_\theta \left( \partial_\theta a_\phi - \partial_\phi a_\theta \right) \, .
\end{align}
Because $g_{\mu\nu}=g^{(0)}_{\mu\nu}$ can be regarded to be constant, we may assume
\begin{align}
\label{phi19}
A_\mu = C_\mu \e^{-i\omega t + i p_r r + i p_\theta \theta + im \phi}\, ,
\end{align}
with constants, $C_\mu$, $\omega$, $p_r$, $p_\theta$ and an integer $m$.
Then Eqs.~(\ref{phi18t}), (\ref{phi18r}), (\ref{phi18theta}), and (\ref{phi18phi}) become algebraic equations,
\begin{align}
\label{phi20t}
0=&\, g^{(0)t\phi} \omega^2 C_\phi - g^{(0)rr} p_r \omega C_r - g^{(0)\theta\theta} p_\theta \omega C_\theta
 - g^{(0)\phi\phi} m \omega C_\phi \, , \\
\label{phi20r0}
0=&\, g^{(0)tt} \omega^2 C_r - g^{(0)t\phi} \omega \left( m C_r - p_r C_\phi \right)
 - g^{(0)\phi t} m \omega C_r \nonumber \\
&\, + g^{(0)\theta\theta} p_\theta \left( p_\theta C_r - p_r C_\theta \right)
+ g^{(0)\phi\phi} m \left( m C_r - p_r C_\phi \right) \, , \\
\label{phi20theta0}
0=&\, g^{(0)tt} \omega^2 C_\theta - g^{(0)t\phi} \omega \left( m C_\theta - p_\theta C_\phi \right) \nonumber \\
&\, - g^{(0)\phi t} m \omega C_\theta + g^{(0)rr} p_r \left( p_r C_\theta - p_\theta C_r \right)
+ g^{(0)\phi\phi} m \left( m C_\theta - p_\theta C_\phi \right) \, , \\
\label{phi20phi0}
0=&\, g^{(0)tt} \omega^2 C_\phi - g^{(0)\phi t} m \omega C_\phi + g^{(0)rr} p_r \left( p_r C_\phi - m C_r \right)
+ g^{(0)\theta\theta} p_\theta \left( p_\theta C_\phi - m C_\theta \right) \, .
\end{align}
By defining
\begin{align}
\label{phi21}
p^2 \equiv g^{(0)tt} \omega^2 - 2 g^{(0)t\phi} \omega m + g^{(0)rr} {p_r}^2 + g^{(0)\theta\theta} {p_\theta}^2 + g^{(0)\phi\phi} m^2\, ,
\end{align}
we find
\begin{align}
\label{phi22t}
0=&\, \omega \left( g^{(0)t\phi} \omega C_\phi - g^{(0)rr} p_r C_r - g^{(0)\theta\theta} p_\theta C_\theta - g^{(0)\phi\phi} m C_\phi \right) \, , \\
\label{phi20r}
0=&\, p^2 C_r + p_r \left( g^{(0)t\phi} \omega C_\phi - g^{(0)rr} p_r C_r - g^{(0)\theta\theta} p_\theta C_\theta - g^{(0)\phi\phi} m C_\phi \right) \, , \\
\label{phi20theta}
0=&\, p^2 C_\theta + p_\theta \left( g^{(0)t\phi} \omega C_\phi - g^{(0)rr} p_r C_r - g^{(0)\theta\theta} p_\theta C_\theta - g^{(0)\phi\phi} m C_\phi \right) \, , \\
\label{phi20phi}
0=&\, p^2 C_\phi + p_\phi \left( g^{(0)t\phi} \omega C_\phi - g^{(0)rr} p_r C_r - g^{(0)\theta\theta} p_\theta C_\theta - g^{(0)\phi\phi} m C_\phi \right) \, .
\end{align}
Eq.~(\ref{phi22t}) gives the Gauss-law constraint,
\begin{align}
\label{phi23t0}
0= g^{(0)t\phi} \omega C_\phi - g^{(0)rr} p_r C_r - g^{(0)\theta\theta} p_\theta C_\theta - g^{(0)\phi\phi} m C_\phi \, ,
\end{align}
and Eqs.~(\ref{phi20r}), (\ref{phi20theta}), and (\ref{phi20phi}) give the following dispersion relation,
\begin{align}
\label{phi24}
0 = p^2 = g^{(0)tt} \omega^2 - 2 g^{(0)t\phi} \omega m + g^{(0)rr} {p_r}^2 + g^{(0)\theta\theta} {p_\theta}^2 + g^{(0)\phi\phi} m^2\, ,
\end{align}
which can be solved with respect to $\omega$
\begin{align}
\label{phi25}
\omega = \omega_\pm \equiv \frac{g^{(0)t\phi}}{g^{(0)tt}} m \pm \sqrt{ \left( \frac{g^{(0)t\phi}}{g^{(0)tt}} \right)^2 m^2
 - \frac{1}{g^{(0)tt}} \left( g^{(0)rr} {p_r}^2 + g^{(0)\theta\theta} {p_\theta}^2 + g^{(0)\phi\phi} m^2 \right) }\, .
\end{align}
For the spherically symmetric case where $g^{(0)t\phi}=0$, $\omega$ is invariant under the change of the signature of $m$, $m\to -m$.
Because Eq.~(\ref{phi23t0}) tells
\begin{align}
\label{phi23t}
C_\phi = \frac{1}{g^{(0)\phi\phi} m } \left( g^{(0)t\phi} \omega C_\phi - g^{(0)rr} p_r C_r - g^{(0)\theta\theta} p_\theta C_\theta \right) \, ,
\end{align}
the signature of $m$ expresses the difference between the left rotating helicity and the right rotating helicity.
Although in any axially symmetric and rotating solution of space-time like the Kerr solution, $g^{(0)t\phi}$ does not vanish, there are several characteristic structures
in our model.
Eq.~(\ref{sol:metric_elements}) tells the explicit form of $g_{t\phi}$.
The first term of $g_{t\phi}$ is common in rotating black hole solutions like the Kerr solution or the Kerr-Newman solution but the second term is
characteristic of our model.
The first term vanishes when $\theta=0$ or $\theta=\pi$, which corresponds to the north pole or the south pole.
This tells that the difference between the dispersion relations for the left and right rotations vanishes at the poles if there is not the second term,
which does not have the $\theta$ dependence.
Furthermore, the second term increases when $H\equiv M^2 - q^2$ goes to vanish, that is, in the extremal limit.
Of course, the second term appears as a correction coming from the Chern-Simons term and therefore the expression is valid only when
the Chern-Simons coupling is small but the term can dominate more than the first term near the limit.
The above characteristic structures will give some effects on the photon which goes through the region near the black hole although it
could be an interesting subject to investigate how the effects on the photon can be observed.

\section{Conclusion and discussions}
\label{conclusions}

A new non-trivial natural slowly rotating BH solution using the dCS gravitational theory has been derived \cite{Yunes:2007ss}.
The charge in the study of BH solutions is important because BH generated in a collider could have an electric field.
Therefore, we included the effect of electric charge in the dCS field equations by including the effect of Maxwell's electromagnetic field.
We are interested in the effects of the dCS term and therefore we did not study the non-dynamical case
because the result of this theory is not changed from the result of the Reissner-Nordstr\" om black hole solution.

If we include the electric charge and the electromagnetic field, the scalar of the SC field in \eqref{eq:constraint} is modified, and therefore,
the expression of the scalar field $\varphi^{(1,0)}$ and $\varphi^{(1,1)}$ are not equivalent to the forms given in \cite{Yunes:2007ss}.
Moreover, when we apply the expression of the CS scalar field $\varphi^{(1,0)}$ in the field equation \eqref{eom}, we obtain the rectification of the metric up to order $\epsilon$.
This rectification $\omega^{(1,1)}$ gives an asymptotic form of order $\mathcal{O}\left(\frac{1}{r^7}\right)$ for large $r$,
which is much weaker than the expression given in \cite{Yunes:2007ss} where the leading order behavior is $\mathcal{O}\left(\frac{1}{r^6}\right)$.
Moreover, we found that the leading form of the Pontryagin density $R_{\nu\mu\alpha\beta} \tilde R^{\mu\nu\alpha\beta}$
has the form $\mathcal{O}\left(\frac{1}{r^7}\right)$ when $r$ is large, which agrees with the results given in the literature, however,
the next-to-leading term is of order $\mathcal{O}\left(\frac{1}{r^8}\right)$, which depends on the charge and is stronger than the one given
in \cite{Yagi:2012ya}.
Moreover, we studied the stability of the charged BH solution using the geodesic deviations. We obtain the condition of the stability and discuss its behavior graphically.

We also investigate the possibility that the BH in this paper could be found by any observation.
We focused on the polarization of the photon which propagates near the black hole.
In the case of the Kerr solution or the Kerr-Newman solution, the difference between the dispersion relations for the left and right rotations
vanishes at the north and south poles but in the BH solution in this paper, the difference does not vanish there.
Furthermore, in the extremal limit. where $H\equiv M^2 - q^2$ goes to vanish, the correction coming from the Chern-Simons term becomes dominant.
Therefore it could be an interesting subject to investigate how these effects on the photon can be observed.

Finally, we close our study with the following:
In the present paper, we derive the charged electric BH solution in dCS by using a linear Maxwell field.
Other forms of the Maxwell field, i.e., non-linear forms of the Maxwell field are not beneficial
because the asymptotic form of the non-linear Maxwell field will be $\mathcal{O}(\epsilon^2)$.
Another case that could be interesting to study is to assume the magnetic field in addition to the electric field.
This study will be carried out elsewhere.
Also, another interesting case is to take into account the effect of the electric charge and the potential $V(\varphi)$ for the scalar field $\varphi$
because we only considered the case $V(\varphi)=0$ in this paper.
Also, this case will be our future work.
\section*{Acknowledgment}
The authors would like to thank the anonymous referee for her/his critical comments which put the paper in clear formalism.

\end{document}